\documentclass[fleqn,usenatbib]{mnras}

\usepackage{newtxtext,newtxmath}
\usepackage[T1]{fontenc}
\usepackage{ae,aecompl}

\usepackage{graphicx}	% Including figure files
\usepackage{amsmath}	% Advanced maths commands
\usepackage{amssymb}	% Extra maths symbols
\usepackage{pdflscape}	% Landscape pages
\usepackage{threeparttable}
\newcommand{\pyan}{\texttt{pyaneti}}
\newcommand{\sme}{\texttt{SME}}
\newcommand{\tess}{\emph{TESS}}
\newcommand{\target}{TOI\,813}
\newcommand{\targetb}{\target\,b}

\newcommand{\vespa}{\texttt{VESPA}}

\newcommand\vsini{$v$\,sin\,$i_\star$}    
 
\newcommand\vmic{$v_{\rm mic}$}
\newcommand\vmac{$v_{\rm mac}$}

\newcommand\teff{$T_{\rm eff}$}
\newcommand\logg{log\,{\it g$_\star$}}

\newcommand{\kms}{\,km\,s$^{-1}$} 
\newcommand{\ms}{\,m\,s$^{-1}$}

\newcommand{\smass}[1][$M_{\odot}$]{ $ 1.32 \pm 0.06 $ #1} 
\newcommand{\sradius}[1][$R_{\odot}$]{ $1.94 \pm 0.10 $ #1}
\newcommand{\stemp}[1][$\mathrm{K}$]{ $ 5907 \pm 150 $ #1 }

\newcommand{\sage}[1][$\mathrm{Gyr}$]{$3.73 \pm 0.62$ #1}

\newcommand{\loggmg}[1][gcc]{ $ 3.86 \pm 0.14 $ #1 }
\newcommand{\loggca}[1][gcc]{ $ 3.85 \pm 0.20 $ #1 }
\newcommand{\vsinival}[1][\kms]{ $ 8.2 \pm 0.9 $ #1 }
\newcommand{\FeH}[1][\dex]{ $ 0.10 \pm 0.10 $ #1 }

\newcommand{\Tzerob}[1][days]   {$1370.7836 _{ - 0.0062 } ^ { + 0.0072 }$~#1} 
\newcommand{\Pb}[1][days]   {$83.8911 _{ - 0.0031 } ^ { + 0.0027 }$~#1}

\newcommand{\bb}[1][ ]   {$0.3 _{ - 0.19 } ^ { + 0.18 }$~#1} 
\newcommand{\arb}[1][ ]   {$47.2 _{ - 3.8 } ^ { + 2.1 }$~#1} 
\newcommand{\rrb}[1][ ]   {$0.03165 _{ - 0.00061 } ^ { + 0.00072 }$~#1} 
\newcommand{\rpb}[1][$R_{\oplus}$]   {$6.71 \pm 0.38 $~#1} 
 
\newcommand{\ib}[1][deg]   {$89.64 _{ - 0.27 } ^ { + 0.24 }$~#1} 
\newcommand{\ab}[1][AU]   {$0.423 _{ - 0.037 } ^ { + 0.031 }$~#1} 

\newcommand{\RMbLC}[1][${\rm m\,s^{-1}}$]   {$7.55 _{ - 0.37 } ^ { + 0.32 }$~#1} 
\newcommand{\insolationb}[1][${\rm F_{\oplus}}$]   {$23.1_{ - 3.1 } ^ { + 4.6 }$~#1} 
\newcommand{\denstrb}[1][${\rm g\,cm^{-3}}$]   {$0.283 _{ - 0.064 } ^ { + 0.039 }$~#1} 
\newcommand{\densspb}[1][${\rm g\,cm^{-3}}$]   {$0.254 _{ - 0.037 } ^ { + 0.046 }$~#1} 
\newcommand{\Teqb}[1][K]   {$610 _{ - 21 } ^ { + 28 }$~#1}

\newcommand{\qoneSC}[1][]   {$0.59 _{ - 0.27 } ^ { + 0.28 }$~#1} 
\newcommand{\qtwoSC}[1][]   {$0.22 _{ - 0.15 } ^ { + 0.25 }$~#1}

\newcommand{\Tzerobone}[1][days]{$1370.800 _{ - 0.011 } ^ { + 0.013 }$~#1}
\newcommand{\Tzerobtwo}[1][days]{$1454.6767 _{ - 0.0081 } ^ { + 0.0077 }$~#1} 
\newcommand{\Tzerobthree}[1][days]{$1538.5625 _{ - 0.0039 } ^ { + 0.0043 }$~#1}  
\newcommand{\Tzerobfour}[1][days]{$1622.4580 _{ - 0.0041 } ^ { + 0.0044 }$~#1}  
\title[\targetb]{Planet Hunters TESS I: \target, a subgiant hosting a transiting Saturn-sized planet on an 84-day orbit}

\author[Eisner et al.]
{N. L. Eisner ,$^{1}$\thanks{E-mail: nora.eisner@new.ox.ac.uk}
O. Barrag\'an,$^{1}$
S. Aigrain,$^{1}$
C. Lintott,$^{1}$
G. Miller,$^{1}$
N. Zicher,$^{1}$
\newauthor
T. S. Boyajian,$^{2}$
C. Brice\~{n}o,$^{3}$
E. M. Bryant,$^{4,5}$ 
J. L. Christiansen,$^{6}$
A. D. Feinstein,$^{7}$
\newauthor
L. M. Flor-Torres,$^{8}$
M. Fridlund,$^{9,10}$
D. Gandolfi,$^{11}$
J. Gilbert,$^{12}$
N. Guerrero,$^{13}$
\newauthor
J. M. Jenkins,$^{6}$
K. Jones, $^{1}$
M. H. Kristiansen,$^{14}$
A. Vanderburg,$^{15}$
N. Law,$^{16}$
\newauthor
A. R. L\'opez-S\'anchez,$^{17,18}$
A. W. Mann,$^{16}$
E. J. Safron,$^{2}$
M. E. Schwamb,$^{19,20}$
\newauthor
K. G. Stassun,$^{21,22}$
H. P. Osborn,$^{23}$
J. Wang,$^{24}$
A. Zic,$^{25,26}$ 
C. Ziegler,$^{27}$
F. Barnet,$^{28}$
\newauthor
S. J. Bean,$^{28}$
D. M. Bundy,$^{28}$
Z. Chetnik,$^{28}$
J. L. Dawson,$^{28}$
J. Garstone,$^{28}$
\newauthor
A. G. Stenner,$^{28}$
M. Huten,$^{28}$
S. Larish,$^{28}$
L. D. Melanson$^{28}$
T. Mitchell,$^{28}$
\newauthor
C. Moore,$^{28}$
K. Peltsch,$^{28}$
D. J. Rogers,$^{28}$
C. Schuster,$^{28}$
D. S. Smith,$^{28}$
\newauthor
D. J. Simister,$^{28}$
C. Tanner,$^{28}$
I. Terentev $^{28}$ 
and A. Tsymbal$^{28}$\\
\\
Affiliations are listed at the end of the paper.
}

\date{Submitted on 12 September 2019; revised on 19 December 2019; accepted for publication by MNRAS on 8 January 2020.}

\pubyear{2019}

\begin{document}
\label{firstpage}
\pagerange{\pageref{firstpage}--\pageref{lastpage}}
\maketitle

\begin{abstract}
We report on the discovery and validation of \targetb\ (\textit{TIC 55525572 b}), a transiting exoplanet identified by citizen scientists in data from  NASA's Transiting Exoplanet Survey Satellite (\tess) and the first planet discovered by the Planet Hunters \tess\ project. The host star is a bright ($V = 10.3$\,mag) subgiant ($R_\star=1.94\,R_\odot$, $M_\star=1.32\,M_\odot$). It was observed almost continuously by \tess\ during its first year of operations, during which time four individual transit events were detected. The candidate passed all the standard light curve-based vetting checks, and ground-based follow-up spectroscopy and speckle imaging enabled us to place an upper limit of $2\,M_{\rm Jup}$ (99\% confidence) on the mass of the companion, and to statistically validate its planetary nature. Detailed modelling of the transits yields a period of \Pb, a planet radius of \rpb\, and a semi major axis of \ab. The planet's orbital period combined with the evolved nature of the host star places this object in a relatively under-explored region of parameter space. We estimate that \targetb\ induces a reflex motion in its host star with a semi-amplitude of $\sim6$\ms, making this a promising system to measure the mass of a relatively long-period transiting planet. \
\end{abstract}


%%%%%%%%%%%%%%%%%%%%%%%%%%%%%%%%%%%%%%%%%%%%%%%%%

%%%%%%%%%%%%%%%%% BODY OF PAPER %%%%%%%%%%%%%%%%%%

\section{Introduction}

The Transiting Exoplanet Survey Satellite \citep[\protect\tess;][]{ricker15} is the first nearly all-sky space-based transit search mission. Over the course of its two year nominal mission, \tess\ will observe 85 per cent of the sky, split up into 26 observational sectors (13 per hemisphere) that extend from the ecliptic pole to near the ecliptic plane. Targets located at low ecliptic latitudes (around 63 per cent of the sky) will be monitored for $\approx$27.4 continuous days, while a total of $\sim$2 per cent of the sky at the ecliptic poles will be observed continuously for $\sim$356 days. This observational strategy means that \tess\ will provide us with a plethora of short period planets ($\lesssim 20$\,d) around bright ($V \lesssim 11$\,mag), nearby stars, which will allow for detailed characterization \citep[e.g.,][]{Barclay2018,Gandolfi2018,Huang2018,Esposito2019}.

Longer-period planets will, however, be significantly more difficult to detect. This is partly because the transiting probability of a planet decreases with increasing orbital distance from the star and partly because detection pipelines typically require two or more transit events in order to gain the signal-to-noise ratio (S/N) needed for detection and often three or more events to confirm a periodicity. The requirement of multiple transit events, in particular, poses a problem for the automated detection of long-period planets in the \tess\ light curves. This is because only the targets close to the poles will be monitored across multiple observational sectors and will thus have light curves with longer observational baselines and the opportunity to detect longer period planets with multiple transit events. This is reflected in the early \tess\ results\footnote{\url{ https://exofop.ipac.caltech.edu/tess/}}: $88\%$ of the first 1075 \tess\ Objects of Interest (TOI) have periods $< 15$ days and $95\%$ have periods $< 30$ days. Non-standard methods, such as machine learning \citep[e.g.,][]{Pearson2018, Zucker2018}, probabilistic transit model comparison \citep[e.g.,][]{Foreman-Mackey2016} or citizen science \citep[e.g.,][]{fischer12}, can in some cases outperform standard transit search pipelines for longer-period planets, and are often sensitive to single transit events that the pipelines routinely ignore. This motivated us to initiate systematic searches for transits in the \tess\ data using some of these alternative methods.

In this paper, we announce the detection and statistical validation of \targetb\ (\textit{TIC 55525572 b}), a Saturn-sized planet orbiting around a bright subgiant star. The candidate was initially identified as a single-transit event by citizen scientists taking part in the Planet Hunters \tess\ (PHT) project. \footnote{\url{www.planethunters.org}} The later detection of further transit events allowed us to constrain the orbital period to $\sim$ 84-days, making it, to the best of our knowledge, the longest-period validated planet found by \tess\ to date. The stellar brightness together with the expected Doppler semi-amplitude of $\sim6$\ms\, make this target one of the few long-period transiting planets for which a precise mass measurement is feasible through radial velocity (RV) observations. 

The host star is a subgiant that is in the process of moving away from the main-sequence and onto the red giant branch. Evolved stars are not normally prime targets for transit surveys, as their large radii make the transits shallower, longer and harder to detect. They also display relatively large projected rotational velocities (owing to their large radii), making precise radial velocity measurements more difficult. Evolved stars are also comparatively scarce in the Solar neighborhood, as the subgiant and giant phases of stellar evolution are short-lived. Consequently, relatively few planets are known around subgiants, yet these offer a unique opportunity to test how a mature planet responds to the increase in stellar flux and proximity as the star expands. This new discovery thus adds to the relatively small but important sample of known planets around evolved stars.

The layout of the remainder of this paper is as follows. We introduce the PHT project in Section~\ref{sec:citizen}, and describe the discovery of the \targetb\ in the \tess\ data in Section~\ref{sec:Discovery}. In Sections~\ref{sec:stpar} and \ref{sec:validation}, we report on the determination of the stellar parameters, and the statistical validation of the planet, respectively. The final planet parameters are derived and discussed in Section~\ref{sec:resdic}, and we present our conclusions in Section~\ref{sec:conclude}.

\section{The citizen science approach}
\label{sec:citizen}

\begin{figure}
    \centering
    \includegraphics[width=0.47\textwidth]{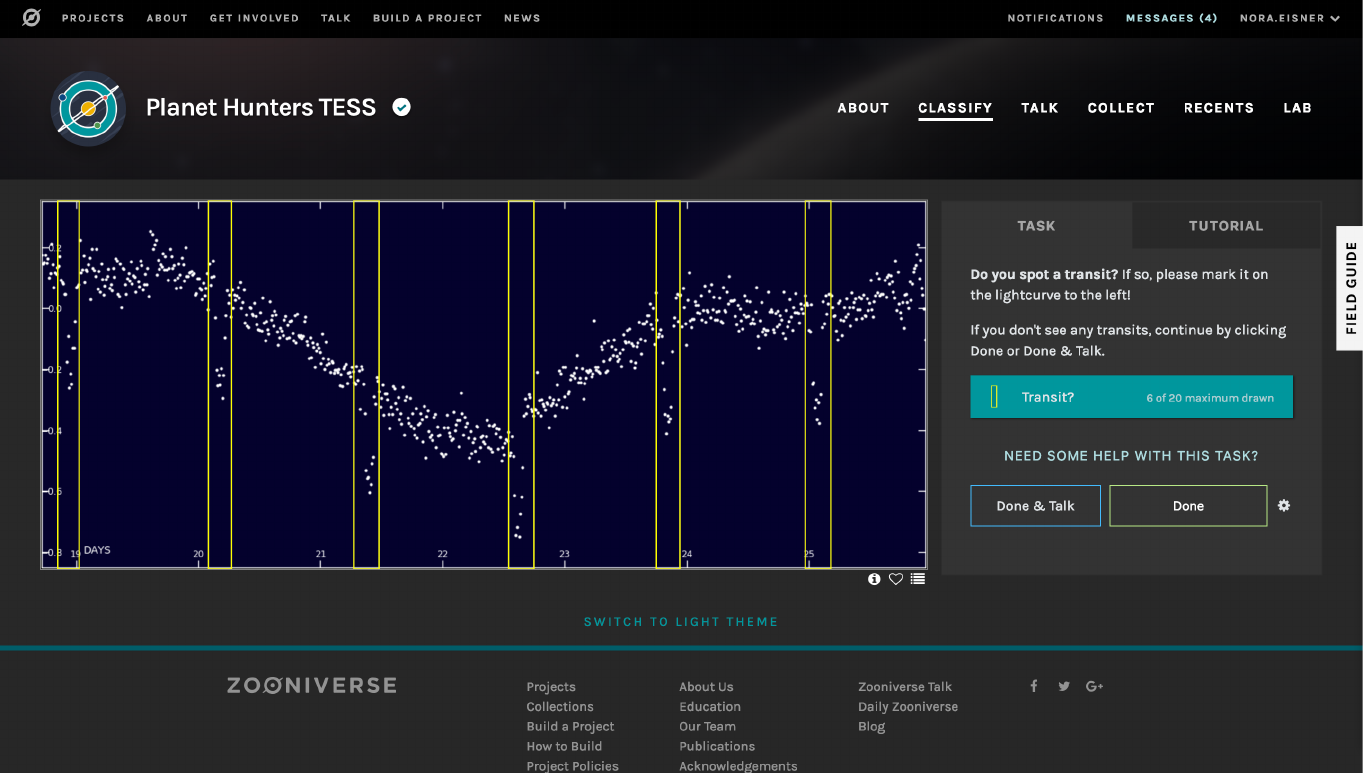}
    \caption{The PHT web interface, as it appeared for Sectors 1--10, with a randomly chosen $\sim$7-day light curve. Volunteers use a mouse drag to register transit-like events, as is shown by the yellow boxes.}
    \label{fig:interface}
\end{figure}

PHT is hosted by Zooniverse, the world's largest and most successful citizen science platform \citep{lintott08, lintott11}. The primary goal of the project is to harness the power of citizen science to find transit events in the \tess\ data that were missed by the main \tess\ pipeline and by other teams of professional astronomers. The project works by displaying \tess\ light curves to volunteers and asking them to mark any transit-like signals by drawing a column over them, as shown in Figure~\ref{fig:interface}. 

PHT builds on the success of the original Planet Hunters project \citep[PH;][]{fischer12}, which used \textit{Kepler} and \textit{K2} data. The initial PH had a detection efficiency $>85\%$ for planets larger than $4\,R_\oplus$ \citep{schwamb12} and detected several noteworthy systems, including the first planet in a quadruple star system \citep{schwamb13} and gas giants orbiting in the habitable zone of their host star \citep{wang13,schmitt14a,schmitt14b}, as well as a large number of candidates that were not found by the main \textit{Kepler} pipeline or other teams \citep{lintott13,wang15}. PH demonstrated that volunteers can outperform automated detection pipelines for certain types of transits, especially single (long-period) transits, as well as aperiodic transits \citep[e.g. circumbinary planets;][]{schwamb13} and planets around rapidly rotating, active stars \citep[e.g., young systems,][]{fischer12}. Additionally, the highly unusual irregular variable KIC\,8462852 was discovered as part of PH \citep{boyajian16}, highlighting the power of citizen science to identify rare but noteworthy objects.

Overall, the bulk of PH's contributions to the \textit{Kepler} findings were long-period planet candidates. In fact, PH found $\sim10\%$ of the \textit{Kepler} candidates with periods $> 100$ days, and $\sim50\%$ of those with periods $> 600$ days (Fischer priv.\ comm.). This is because the PH volunteers typically identify transit-like events one by one, rather than by exploiting their periodic nature as most automated transit-search algorithms do. They are thus equally likely to find a planet candidate that produces only one transit event in a given light curve as they are to find multiple transit events, as was shown by \citet{schwamb12}. 

%The fact that PH was able to detect planets that were otherwise missed testifies to the remarkable pattern recognition abilities of its human participants.

The importance of citizen science in the field of planet detection was also demonstrated by the highly successful Exoplanet Explorer \citep{Christiansen2018} Zooniverse project, which used \textit{K2} data. With the help of volunteers the project has so far led to the validation of multiple interesting planetary systems \citep[e.g., ][]{Feinstein2019,Christiansen2018} as well as many new candidates \citep[e.g., ][]{Zink2019}.

The PHT project was launched with the first public \tess\ data release in December 2018, and volunteers have since classified every two-minute cadence light curve from each Sector, typically within two weeks of that sector's release. By 9 September 2019, the PHT volunteers had classified almost 250\,000 $\sim$30-day light curves. Until Sector 10, the PHT interface was extremely similar to PH: each light curve was split into typically four 7-day segments, and pre-generated plots of the light curve, binned to 14-minute sampling, were uploaded to the website and shown to the volunteers. As of Sector 11, the PHT interface displays the entire $\sim$27-day light curves binned to 10-minute sampling, and the project has the added capability to zoom in on the data.

%As of Sector 10, we have introduced a new interactive interface, which displays the entire \tess\ light curves at 10-minute sampling, and enables the user to zoom in and out.

Volunteers are also shown simulated light curves where we have injected transit like signals into light curves, resulting in a S/N of at least 7. These allow us to evaluate the sensitivity of the project and assess the skill of each individual volunteer. Each real light curve (or light curve segment) is seen by 8 to 15 volunteers and the significance of each transit-like event is evaluated based on all the marked transits \citep[a similar algorithm is described in][]{schwamb12}. Volunteers are also given the option to discuss their findings with each other, as well as with the science team, via the \textit{Talk} discussion forum. \footnote{\url{ www.zooniverse.org/projects/nora-dot-eisner/planet-hunters-TESS/talk}}

PHT engages a very large number of members of the public, some of them over a considerable time period. There are $>11,000$ registered participants, and many more who are not registered. Some spend only a few minutes on the site, others regularly devote several hours per week to the project. While most volunteers simply mark transit-like events, a significant proportion go much further, downloading and analysing \tess\ light curves at their own initiative.

\section{Discovery of \protect\targetb\ in the \protect\tess\ data} \label{sec:Discovery}

\subsection{\protect\tess\ data} \label{subsec:tess}

\target\ \citep[TIC\,55525572;][]{Stassun19} is located at high ecliptic latitude and was observed almost continually by \tess\ during its first year of observations, from Sectors 1--13 except in Sector 7,  according to the Web TESS Viewing Tool (WTV)\footnote{\url{https://heasarc.gsfc.nasa.gov/cgi-bin/tess/webtess/wtv.py}}. However, it was only included in the list of targets for which two-minute cadence observations are available from Sector 4 onwards. Prior to that, i.e.\ during Sectors 1--3, it was included in the Full Frame Images (FFIs) collected every 30 min only. 

Only the two-minute cadence targets are searched by PHT, which uses the Pre-Search Data Conditioning (PDC-MAP) light curves produced by the \tess\ pipeline at the Science Processing Operations Center \citep[SPOC; ][]{jenkins16}. These light curves have been corrected for both known pixel-level instrumental effects and systematics common to many light curves. We downloaded the light curves from the Mikulski Archive for Space Telescopes (MAST)\footnote{\url{http://archive.stsci.edu/tess/}}, and discarded observations which were flagged by the SPOC pipeline as affected by various instrumental anomalies. 

After the initial detection of the transits (see below), we used the FFIs for Sectors 1--3 to produce light curves for \target\ using the open source package \texttt{eleanor} \citep[v0.1.8,][]{Feinstein2019}, which performs background subtraction, aperture photometry, and detrending for any source observed in the FFIs. The extracted FFI light curves were corrected for jitter by quadratically regressing with centroid position. 

%\textcolor{red}{Include here any relevant details -- anything the reader would need to reproduce the light curve extraction. Does eleanor do any systematics removal or detrending?}.

At the time of writing, the \tess\ data were publicly available up to, and including, Sector 13. The detailed analysis of \targetb's light curve, including the vetting checks described in Section~\ref{subsec:vetting} and the transit modelling (Section~\ref{subsec:modelling}) was carried out using the two-minute cadence Light Curve (LC) files and Target Pixel Files (TPFs) produced by versions 3.3.51 to 3.3.75 of the SPOC pipeline. 

In total, the light curve of \target\ consists of 149265 flux measurements between barycentric \tess\ Julian Date (BTJD, defined as BJD-2457000) 1354.13650 and 1682.35665. The 2-minute cadence light curves have a median flux of 18338 counts/sec and a typical RMS scatter of 122 counts/sec.
%(\textcolor{red}{Note to self: this is without detrending - add with detrending too?}

% TESS Julian Date (TJD), defined as BTJD−2457000

\begin{figure*}
    \centering
    \includegraphics[width=\textwidth]{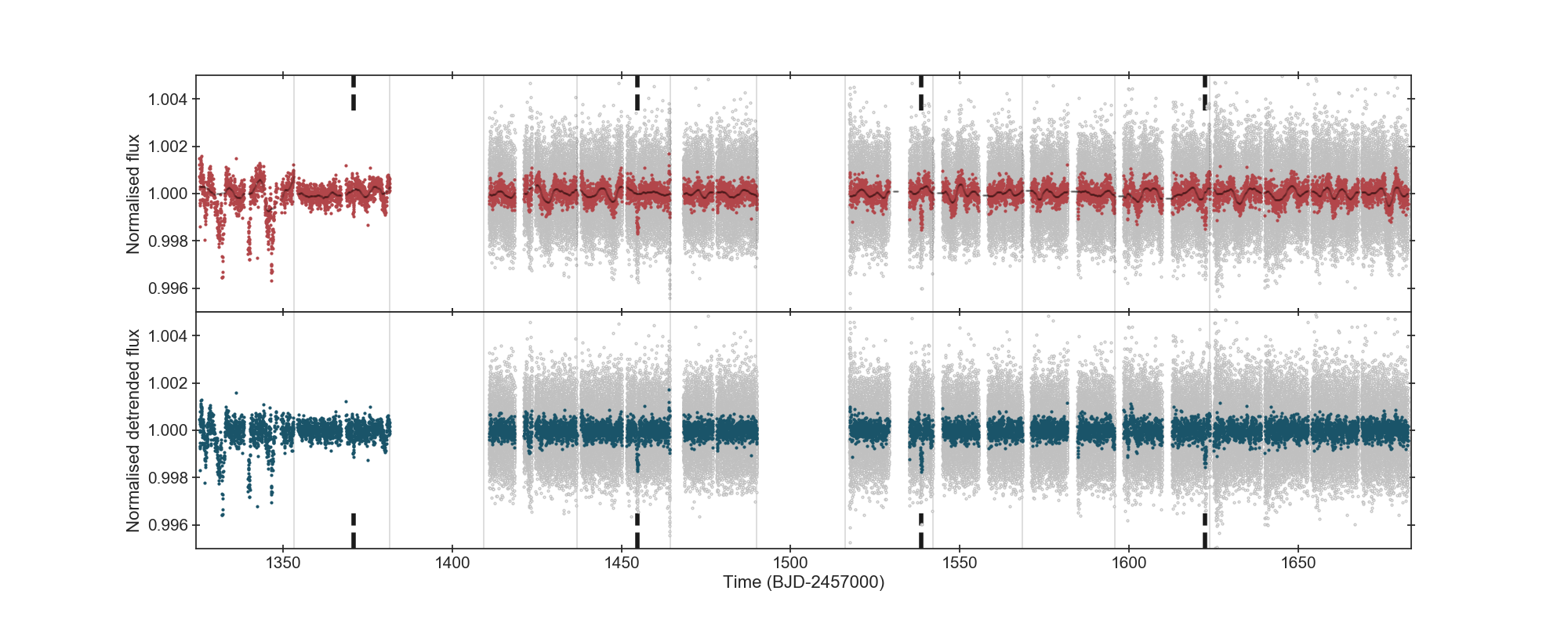}
    \caption{\protect\tess\ light curve for \protect\target\ for Sectors 1--13, except Sector 7. The top panel shows the PDC light curve at a two minute sampling (grey) and a 30-minute sampling (dark red). The black points shows the best fit model used for detrending. The bottom panel shows the detrended light curve, used in the BLS search (see Section~\ref{sec:injectiontest}), also at a 2-minute (grey) and 30-minute (teal blue) sampling. The times of transits are indicated by the short black dashed vertical lines and the end of each sector is depicted by a solid light grey line.}
    \label{fig:full_LC}
\end{figure*}

%\protect\textcolor{red}{I suggest modifying this figure show the FFI data for sectors 1--3 alongside the two-minute cadence data for the remaining sectors. This would work better if the binned version of the short-cadence light curve (dark red points) was binned to 30min to match the cadence of the FFI data. Also I suggest making the lines marking the location of the transits more prominent (larger line width?) and those marking the sector edges less prominent (lower alpha?). Additionally, you might want to show on this figure the smoothed version of the light curve that you compute and subtract before running the BLS transit search -- and perhaps even add a separate panel below with the detrended light curve used in the BLS search (unless you decide to show the light curve after detrending in a separate figure in the "search for additional transits" section"). Finally, don't forget to include Sector 12 (and 13 if released) here.}

\subsection{Discovery of \protect\targetb}

\begin{figure}
    \centering
    \includegraphics[width=0.48\textwidth]{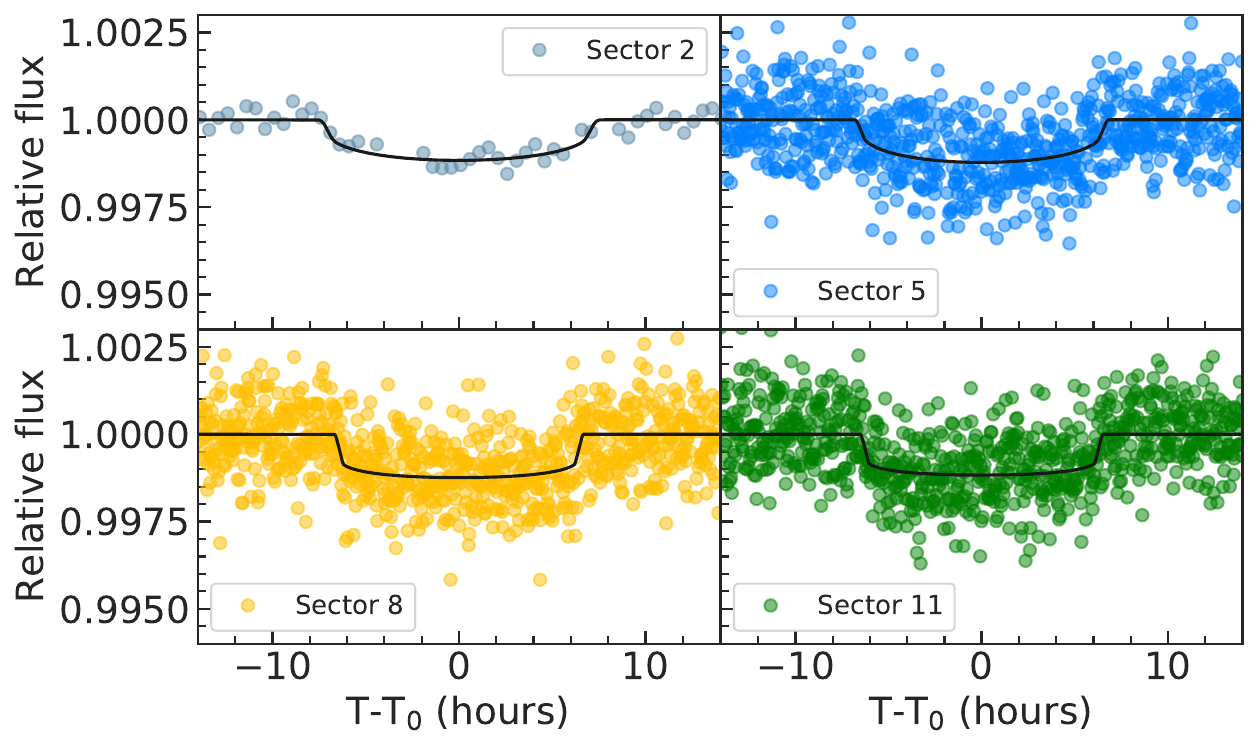}
    \caption{
    Individual transits of \protect\targetb\ in the \protect\tess\ data in Sectors 2 (top left), 5 (top right), 8 (low left) and 11 (low right). The Sector 2 light curve was extracted from the FFIs whereas the other sectors are two-minute cadence observations.} 
    \label{fig:transits_individual}
\end{figure}

Adopting $\sim84$-days as the orbital period of the companion, an additional transit was predicted to have taken place in Sector 2. The target was not included in PHT in Sector 2, as there was no two-minute cadence light curve. We thus extracted the Sector 2 light curve from the FFIs and found the transit by visual inspection at the expected time. A fourth transit was predicted to occur in Sector 11 data of the \tess\ primary Southern Hemisphere survey, and was indeed found in the Sector 11 light curve once that data was released. The full light curve for \target\ is shown in Figure~\ref{fig:full_LC} and the individual transits in Figure~\ref{fig:transits_individual}. Once the PHT team had completed basic vetting tests, we reported this candidate on the ExoFOP website\footnote{\url{ https://exofop.ipac.caltech.edu/tess/index.php}} as a community TESS Object of Interest (cTOI), and it was allocated TOI number 813.01.

%At that point, on 15 May 2019, we submitted it to the Exoplanet Follow-up Observing Program for TESS (ExoFOP-TESS) website\footnote{\url{ https://exofop.ipac.caltech.edu/tess/index.php}} as a cTOI to encourage astronomers around the world to contribute to the follow-up effort.

At the time of the PHT discovery, 27 April 2019, \target\ was not listed as a TOI, nor did it have any Threshold Crossing Events (TCEs). In other words, it was not detected by either the SPOC transit search pipeline, or the Quick Look Pipeline \citep[][; Huang et al. (in prep)]{Fausnaugh18}, which is used by the TESS Science Office (TSO). This is because the planet never exhibits more than one transit in a given sector and the SPOC pipeline requires at least 2 transits for a detection. At the time, the SPOC pipeline had not yet been run on multi-sector data. The QLP light curve from Sector 8 did not advance past initial triage for vetting by the TOI team.

However, concurrently to the citizen science discovery, a different subset of the co-authors of the present paper identified \targetb\ independently, via a manual survey using the LcTools software \citep{Kipping2015}, following the method described by \cite{Rappaport2019}.

The TOI team released a TCE on TIC 55525572 as TIC 55525572.01 on 21 June 2019. The target appeared as a TCE in the SPOC multi-sector planet search in \tess\ sectors 1-9. \footnote{\url{https://archive.stsci.edu/missions/tess/doc/tess_drn/tess_multisector_01_09_drn15_v03.pdf}} The target was initially ranked as low priority for manual vetting, and then classified as a potential planet candidate in group vetting. The first transit of the object occurred around the time of a spacecraft momentum dump, but the second was marked as a potential single transit of a planet candidate. Furthermore, the SPOC pipeline detected the transits of TOI-813 in sectors 5, 8 and 11 in the multisector transit search of sectors 1-13 performed on 7 August 2019 with orbital parameters and planet radius consistent with this work. \footnote{\url{https://mast.stsci.edu/api/v0.1/Download/file/?uri=mast:TESS/product/tess2018206190142-s0001-s0013-0000000055525572-00226_dvr.pdf}}

\subsection{Light curve based vetting checks}
\label{subsec:vetting}

We carried out a number of vetting tests on the \tess\ data, similar to the `Data Validation' step of the SPOC pipeline. These are intended to rule out as many as possible of the false positive scenarios that could have given rise to the detection, whether they are of instrumental or astrophysical origin. 

First, we checked for instrumental false alarms by comparing the light curve around the time of each transit to the star's centroid position and the background flux, which are provided in the light curve files, and to the times of reaction wheel momentum dumps, which occur every 2 to 2.5 days and typically last around half an hour. While observations taken during a momentum dump are flagged and were not used in this work, the satellite pointing remains affected for several hours after each dump, and can result in spurious flux variations due to aperture losses or inter-/intra-pixel sensitivity variations. Additionally, enhanced scattered light in the telescope optics can cause dramatic increases in the  background flux when the Earth, Moon or other solar system planet pass below $25 ^\circ$ from the boresight of any of the cameras. Both of these effects can induce spurious transit-like events. Even though one of the transits of \targetb\ was found to occur at the time of a momentum dump, the other three do not coincide with times where the light curve was potentially affected by enhanced pointing jitter or background flux, and due to the periodicity of all of the transit-events, we believe them all to be real. 

In order to identify other sections of the light curves potentially affected by residual systematics common to the light curves of many different targets, we plotted a histogram of all of the volunteer markings across all subject light curves for each sector. If volunteers tend to mark transit-like events at the same time across many sources, it is likely that these are caused by systematics. We found that none of the transits of \targetb\ observed in Sectors 5, 8 or 11 coincide with a time when volunteers marked an unusually high number of other targets. Finally, we inspected the light curves of all \tess\ two-minute target stars within $0.5 ^\circ$ of \target\ in order to check whether any of them showed flux dips at the times of the transits, finding none. This allowed us to rule out large scale detector anomalies or contamination by a bright, nearby eclipsing binary as the cause for the transit-like signals.

\begin{figure}
    \centering
    \includegraphics[width=0.49\textwidth]{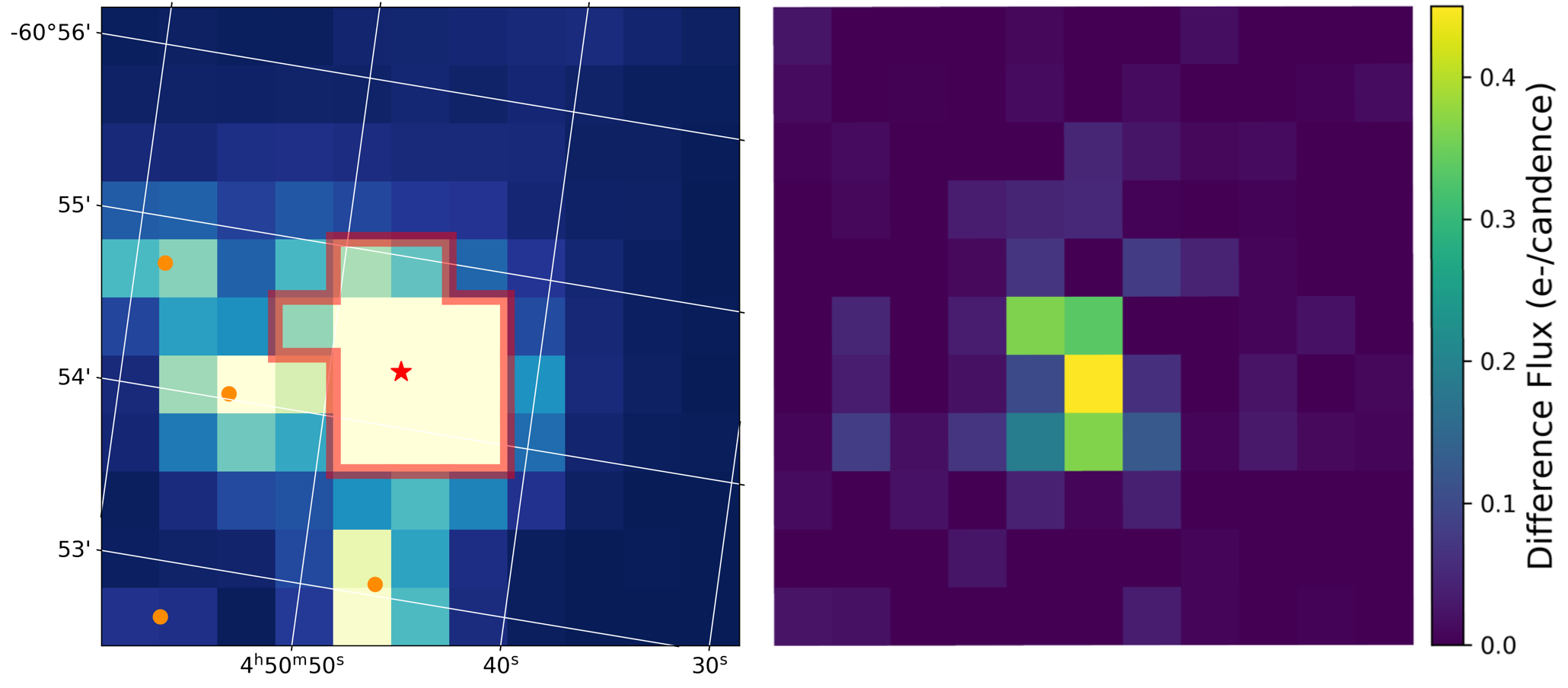}
    \caption{Left: \protect\tess\ image in the vicinity of \target\ throughout Sector 5. The red outline shows aperture mask used to extract the light curve. Neighbouring stars  brighter than V = 16 within 180 arcminutes of \target\ are depicted by orange dots. Right: Difference in/out of transit image during Sector 5.}
    \label{fig:starmap}
\end{figure}

We then carried out a further set of vetting tests aiming to exclude astrophysical false positives. In particular, \tess's large pixel size (21\,arcsec) means that many fainter stars contribute to the flux recorded in the aperture of each target. If any of these faint neighbour is an eclipsing binary, its diluted eclipses can mimic a transit on the main target star. A large fraction of these `blend' scenarios can be ruled out as follows:
\begin{description}
    \item[\textit{Odd and even transit comparison}] We compared the depths, duration and shape of the odd- and even-numbered transits. Slight differences in these would indicate that the `transits' are caused by a near-equal mass eclipsing binary. The phase-folded light curves for the odd- and even-numbered transits were modelled separately (as described in Section~\ref{subsec:modelling}) and were found to have depths and widths consistent with one another to within $0.5 \sigma$. 
  
    \item[\textit{Secondary eclipse search}] We searched for a secondary eclipse (or occultation), which might indicate that the transits are caused by a small but self-luminous companion, such as a brown dwarf or low-mass star.This was done by conducting a grid search for the deepest out-of-transit signal in the phase-folded light curve \citep[see e.g.,][]{rowe15}. This resulted in an upper limit on the depth of any secondary eclipse at the level of $346$ parts per million (ppm).

    \item [\textit{Pixel-level centroid analysis}] A blend with a background eclipsing binary would result in a change in the spatial distribution of the flux within the target aperture during the transits. We checked for this in two ways. First, we computed a difference image by subtracting the in-transit pixel flux from the image obtained during a similar time period immediately before and after each transit. This was done using the Target Pixel Files (TPFs) released alongside the two-minute cadence light curves, after correcting them for systematic effects using Principal Component Analysis \citep[e.g.,][]{Stumpe2012, Smith2012}. The difference image shows only one source, whose location coincides with that of the main target, as shown for the Sector 5 transit in right hand panel of Figure~\ref{fig:starmap}.
    Second, we compared the observed position of the target during, and immediately before and after each transit but no significant differences were found. To quantify this, we used a two-sided Kolmogorov-Smirnov test to see whether the flux-weighted centroid positions in- and out-of-transit are drawn from the same distribution or not. This non-parametric test showed that for all of the transits the detrended x- and y-centroid positions did not differ significantly in-transit compared to out-of-transit, with p-values ranging from 0.34 to 0.98 (where statistically different is defined as p < 0.05).

    \item [\textit{Light curve extraction with different aperture sizes}] Another signature of a blended eclipsing binary would be a change in the depth of the transit depending on the size of the photometric aperture used. To check for this, we extracted the 2-minute cadence light curve with different photometric apertures (by growing or shrinking the default aperture mask shown in the left hand panel of Figure~\ref{fig:starmap} by one pixel). For the 30-minute cadence data, we extracted the light curves with different aperture sizes using eleanor. This analysis, which was carried out for each transit event independently, showed that the aperture size had a minimal effect on the depth and shape of the transit.

    \item [\textit{Nearby companion stars}] We searched for evidence of nearby stars by querying all entries in the Gaia Data Release 2 catalog \citep{gaiadr2} within 180 arcseconds of \target. We found there to be 4 stars within this radius with magnitudes brighter than V=16, as shown in Figure~\ref{fig:starmap}). The SPOC pipeline's difference imaging analysis, however, showed that none of these stars are located within the 3-sigma confusion region around the target. Nonetheless, in order to rule out the closest companion stars we calculated the magnitude difference between \target\ and the faintest companion star that could plausibly cause the observed a transit depth and shape. We used the equations presented in \cite{Vanderburg2019} and the transit parameters derived using an MCMC approach (see Section~\ref{subsec:modelling}) and found the maximum magnitude difference to be 0.4 mag. This allows us to confidently rule out the 14.636 magnitude star located at an angular separation of $\sim$ 19 arcsecond as the cause of the transit signal. Furthermore, the SPOC pipeline \citep{jenkins16} accounts for the contamination of the aperture by the neighbouring stars, so that we do not need to correct the measured transit depth for the effect of this light.

    % To do that, I'd recommend taking your transit fit MCMC chains (which it appears you fit without without any constraint on the stellar density from spectroscopy, right?), and from them calculating the transit duration and the duration of transit ingress/egress. Then, once you have those, you can calculate the faintest star that could plausibly create a transit with the shape that you measured using the following equation. I think we can rule out stars 1.7 or more mags fainter than the target star, which easily shows that the companion 5 mags away at 19 arcseconds is not the source of the transits. You'd want to do this more rigorously with MCMC results, especially fofr t12, which is going to be noisy, but I think it's a way to definitively rule out any other stars and satisfy the referee. 

    %As we do not see a centroid shift during the time of the transit events, we can rule these out as the cause for the dips in light curve of \target. 

    %https://arxiv.org/pdf/1904.04980.pdf   -- difference imaging statistically
    % do we need to quantify this? https://iopscience.iop.org/article/10.3847/0004-637X/822/2/86/meta, https://arxiv.org/pdf/1904.04980.pdf

\end{description}

The above tests rule out many, but not all, of the plausible astrophysical false positive scenarios. While we can state with confidence that none of the nearby stars brighter than $V=16$ (shown in Figure~\ref{fig:starmap}) are the source of the transits, fainter contaminant located nearer the main target cannot be ruled out at this stage. Nonetheless, these vetting tests increased our confidence in the planetary nature of the companion sufficiently to motivate ground-based follow-up observations. 

We also note that \targetb\ independently passed all of the SPOC pipeline Data Validation diagnostic tests in the multisector 1--13 run. These tests included the odd and even transit comparison test, the weak secondary eclipse search test, the ghost diagnostic test (which is similar to this work's increasing aperture size test), and the statistical bootstrap test, which looks at the propensity of other transit-like features in the light curve to cause happenstance false positives \citep{Twicken2018, Li2019}.

\subsection{Search for additional planets}
\label{sec:injectiontest}

The discovery of multiple planets within one system can significantly increase confidence in the planetary nature of a transit-like event. No additional planet candidates were identified by the citizen science campaign, however, the volunteers' detection efficiency is known to decrease notably as a function of planet radius, as was shown by the initial PH project \citep{schwamb12}. We therefore carried out a search for additional transits in the full light curve using the Box Least Squares \citep[BLS; ][]{bls2002} algorithm, after masking the transits of \targetb. Before running the BLS, we used an iterative non-linear filter \citep{Aigrain04} to estimate and subtract residual systematics on timescales $>1.7$ days (see Figure~\ref{fig:full_LC}). The BLS search was carried out on an evenly sampled frequency grid ranging from 0.01 to 1 d$^{-1}$ (1 to 90 days). We used the ratio of the highest peak in the SNR periodogram relative to its standard deviation, known as the signal detection efficiency (SDE) to quantify the significance of the detection. The algorithm found no additional signals with SDE $>$ 7.2 (compared to SDE$\sim$22.4 for the transit of \targetb).

%The algorithm recovered a signal with a period of $\sim$ 13.4-d with an SDE of 7.2 \textcolor{green}{(compared to SDE~22.4 \targetb)}. This period corresponds to the duration of a \tess\ orbit, and therefore is explained by systematical effects due to additional scatter in the light curve around the time when the data is sent back to Earth. We masked these times of increased scatter and found there to be no signals with SDE > \textcolor{red}{XXX}. 

In addition, we searched for further companions by manually inspecting the light curve with LcTools, after having binned the data at 6 points per hour in order to visually enhance undetected signals with low SNRs. No additional transits were identified with either method.

\begin{figure}
    \centering
    \includegraphics[width=0.45\textwidth]{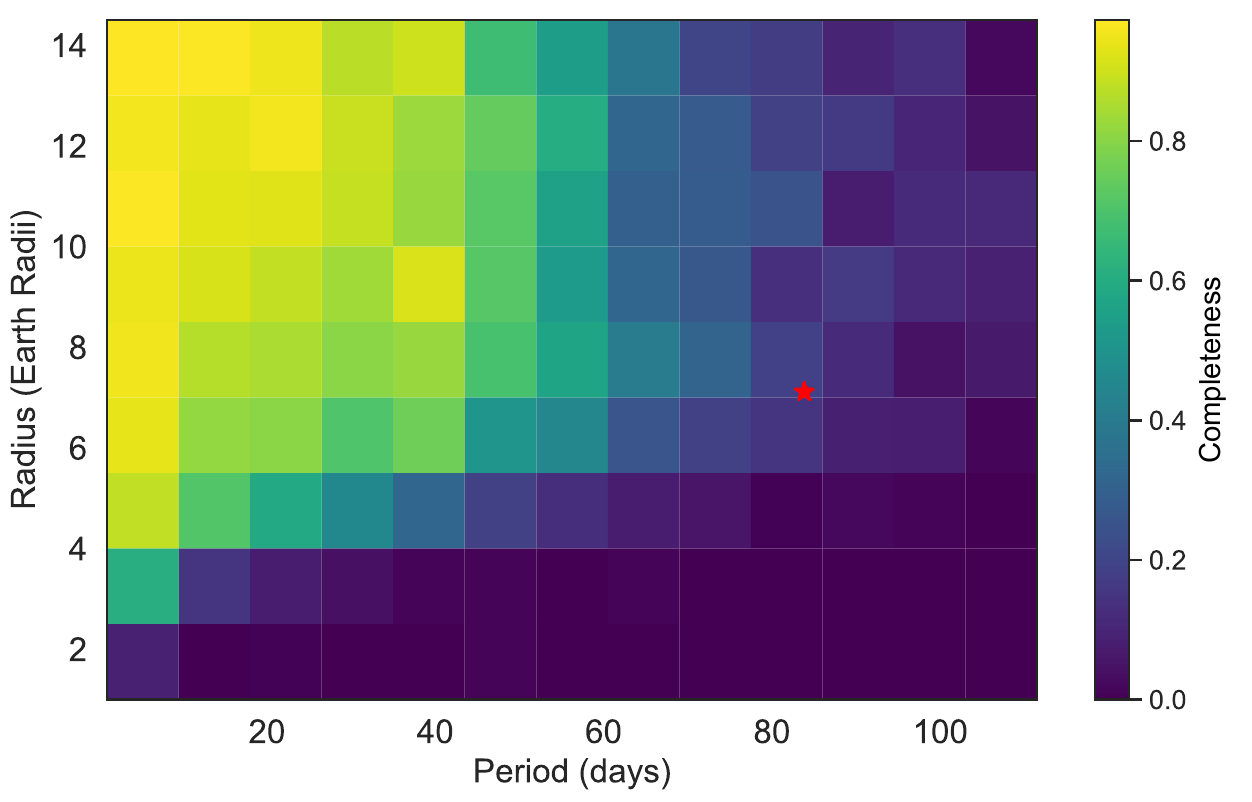} 
    \caption{The recovery completeness of injected transit signals into the light curve of \target\ as a function of the radius vs orbital period, where the signals were recovered using a BLS search. The properties of \protect\targetb\ are depicted by the red star. }
    \label{fig:BLS}
\end{figure}

The non-detection of additional transits suggests that if there are other planets in orbits interior to \targetb, their orbits are either inclined to not transit or are shallower than the TESS detection threshold. 

We used injection and recovery tests to quantify the detectability of additional planets in the \tess\ light curve of \target, injecting artificial signals into the PDC light curve before repeating the masking of \targetb's transit, the detrending and the BLS search. The injected transits were generated using the open source {\tt batman} package \citep{Kreidberg15}, with planet radii and orbital periods sampled at random from logarithmic distributions ranging from 1 to $7.11\,R_\oplus$ and from 0 to 120 days, respectively. The upper bound of the planet radius distribution was chosen to coincide with our best-fit radius for \targetb. For simplicity, the impact parameter and eccentricity were assumed to be zero throughout. We used a quadratic limb-darkening law with $q_1$ and $q_2$ of 0.59 and 0.22, respectively (See Table~\ref{tab:parstarget}).

%\textcolor{red}{You should sample impact parameter uniformly between 0 and 1 (that is what you expect for random inclinations on the sky (NORA's question: in the K2SC paper it's 0?)). For injection tests like this it's fine to assume 0 eccentricity. You should state what you used for the limb-darkening law and parameters (not that it matters, but for completeness\ldots}. 

We simulated and injected transits for 50\,000 planets and attempted to detect them using the BLS algorithm as was done when searching for real signals. For each simulation we identified the highest peak in the BLS periodogram and recorded the corresponding period and orbital phase. The injected signal was considered to be correctly identified if the recovered period and orbital phase were within 1\% of the injected values. We then evaluated the completeness of our search for additional transits by computing the fraction of injected transits that were correctly identified over a grid of period and radius. As we injected the simulated transits into the PDC light curve, the derived detection limits do not take into account the impact of the PDC-MAP systematics correction on transit events, and should therefore be considered optimistic.

The results are shown in Figure~\ref{fig:BLS}. We recover more than $80\%$ of the simulated planets larger than $\sim 5 \,R_{\odot}$ with periods less than $30$ days, dropping to 50\% for periods less than $60$\ days.  Therefore, we cannot rule out the presence of additional planets transiting inside the orbit of \targetb\ at high confidence, particularly sub-Neptunes. It is also interesting to note that the completeness for the period and radius of \targetb\ itself is rather low ($\sim 30\%$): even with multiple transits, such long-period planets are relatively hard to detect in \tess\ data using standard algorithms. Nonetheless, these single transit events are often visible by the human eye, thus highlighting the importance of citizen science.

\section{Refining the stellar parameters} \label{sec:stpar}

\begin{table*}
\centering
  \caption{Stellar parameters of \target. \label{tab:star}}  
  \begin{tabular}{lcc}
  \hline
  Parameter & Value & Source  \\
  \hline
  \multicolumn{3}{l}{\bf Identifiers} \\

  TIC  & 55525572 & \cite{Stassun19} \\
  Gaia DR2 & 4665704096987467776 & \textit{Gaia} DR2$^{(\mathrm{a})}$ \\
  2MASS & J04504658-6054196 & 2MASS$^{(\mathrm{b})}$ \\
  \hline
  
  \multicolumn{3}{l}{\bf Astrometry} \\
  $\alpha_{\rm J2000}$ & 04:50:46.57  &  \textit{Gaia} DR2$^{(\mathrm{a})}$ \\
  $\delta_{\rm J2000}$ & -60:54:19.62  & \textit{Gaia} DR2$^{(\mathrm{a})}$ \\
  Distance (pc) & 265.1535 $\pm$1.582 & \textit{Gaia} DR2$^{(\mathrm{a})}$ \\
  $\pi$ (mas) &  3.7714 $\pm$0.0225 & \textit{Gaia} DR2$^{(\mathrm{a})}$ \\
  
  \hline
  \multicolumn{3}{l}{\bf Photometry} \\
  $B_{T}$ &  $11.189 \pm 0.056$  & \textit{Tycho-2}$^{(\mathrm{c})}$\\ 
  $V_{T}$ &  $10.435 \pm 0.056$  & \textit{Tycho-2}$^{(\mathrm{c})}$ \\ 
  
  $B$ &  $10.905 \pm 0.026$  & APASS$^{(\mathrm{d})}$\\ 
  $V$ &  $10.322 \pm 0.014$  & APASS$^{(\mathrm{d})}$ \\ 
  $g$ &  $10.587 \pm 0.015$  & APASS$^{(\mathrm{d})}$ \\ 
  $r$ &  $10.240 \pm 0.054$  & APASS$^{(\mathrm{d})}$ \\ 
  $i$ &  $10.094 \pm 0.044$  & APASS$^{(\mathrm{d})}$ \\ 
  
  $G$ &  $10.2352 \pm 0.0004$  & \textit{Gaia} DR2$^{(\mathrm{a})}$ \\ 
  
  $J$ &  $9.326 \pm 0.021$  & 2MASS$^{(\mathrm{b})}$ \\ 
  $H$ &  $9.053\pm 0.018$  & 2MASS$^{(\mathrm{b})}$ \\ 
  $K_{s}$  &  $9.029 \pm 0.019$  & 2MASS$^{(\mathrm{b})}$ \\ 
  
  $W1$ (3.35 $\mu$m)  &  $8.992\pm 0.023$  & WISE$^{(\mathrm{e})}$ \\ 
  $W1$ (4.6 $\mu$m) &  $9.027\pm 0.020$  & WISE$^{(\mathrm{e})}$ \\   
  $W1$ (11.6 $\mu$m) &  $8.971\pm 0.022$  & WISE$^{(\mathrm{e})}$ \\ 
  $W1$ (22.1 $\mu$m) &  $9.126\pm 0.271$  & WISE$^{(\mathrm{e})}$ \\   
  
  \hline
  \multicolumn{3}{l}{\bf Kinematics} \\

  \hline
  \multicolumn{3}{l}{\bf Physical Properties} \\
    Stellar mass $M_{\star}$ ($M_\odot$)  &  \smass[] & This work \\
    Stellar radius $R_{\star}$ ($R_\odot$)  & \sradius[]  & This work \\
    \vsini (\kms)  & \vsinival[] & This work  \\
    Stellar density $\rho_\star$ (g\,cm$^{-3}$)  &  \densspb[]  & This work \\   
    Effective Temperature $\mathrm{T_{eff}}$ (K)  & \stemp[] & This work \\
    Surface gravity $\log g_\star$ from Mg I (gcc)  & \loggmg[]  & This work\\ 
    Surface gravity $\log g_\star$ from Ca I (gcc)  & \loggca[]  & This work\\ 
    Iron abundance [Fe/H] (dex)  & \FeH[]  & This work \\
    Star age (Gyr)  & \sage[]   & This work \\
    Spectral Type &  G0 IV   & \citet[][]{Pecaut2013} \\
    $v_{\rm mic}$ (\kms) & 4.4 & \cite{Doyle2014} \\
    $v_{\rm mac}$ (\kms) & 1.21 & \cite{Bruntt2010} \\
    \hline
    %\cite{gaiadr2}\footnote{footnote text}
    %\cite{APASS2014}\footnote{footnote text}  
    %\footnotetext[1]{\cite{gaiadr2}} 
    %\footnotetext[2]{\cite{APASS2014}}
  \end{tabular}
    \begin{tablenotes}\footnotesize
  \item \emph{Note} -- $^{(\mathrm{a})}$ \textit{Gaia} Data Release 2 \citep[DR2; ][]{Gaia2018}. $^{(\mathrm{b})}$ Two-micron All Sky Survey \citep[2MASS; ][]{2MASS2003}. $^{(\mathrm{c})}$ \textit{Tycho}-2 catalog \citep{Hog2000}. $^{(\mathrm{d})}$ AAVSO Photometric All-Sky Survey \citep[APASS; ][]{Munari2014}. $^{(\mathrm{e})}$ Wide-field Infrared Survey Explorer catalog \citep[WISE; ][]{Cutri2013}
\end{tablenotes}
\end{table*}

The stellar parameters provided in the TIC \citep{Stassun19} are based on broad-band photometry, and their precision is therefore limited. We thus acquired moderate, then high-dispersion spectra of \target\ to refine our estimate of the host star's parameters.

\subsection{Spectroscopy} \label{subsec:spec}

We first obtained a moderate spectral resolution ($R=7000$) spectrum of \target\ with the Wide Field Spectrograph instrument on the Australian National University (ANU) 2.3-m telescope \citep{Dopita2007} on the night of 30 April 2019. We obtained three 120 second exposures with the U7000 (S/N $\sim$400) and R7000 (S/N $\sim$700) gratings and the data were reduced using the pyWiFeS data reduction pipeline version 0.7.4 \citep{Childress2014}. 

%This reconnaissance spectrum confirmed the evolved nature of the target and allowed us to run a preliminary statistical validation (see Section~\ref{sec:validation}).

%\textcolor{red}{@OSCAR: Give details of grating, exposure time, SNR, and data reduction and spectral extraction steps. Ideally would also include plot of spectrum, and if possible that plot would also show some of the models used to derive the stellar parameters.} 

%
%With the use of the parallax measurements from Gaia DR2 we %found the distance to the host star to be
%
%parallax = 3.7714 +/- 0.0225 mas
%         = 0.0037714 +/- 0.0000225 as
%         
%distance = 0.265 +/- 0.0016 Mpc
%distance = 265.154 +/- 1.6 pc
%
%Properties of the host star were further constrained to be...

We then acquired a high-resolution ($R \approx 115000$) spectrum with the High Accuracy Radial velocity Planet Searcher \citep[HARPS;][]{Mayor2003} spectrograph on the ESO 3.6-m telescope at La Silla observatory (Chile). The observations were carried out on 14 July 2019 as part of observing program 1102.C-0923. We used an exposure time of 1800 sec leading to a S/N per pixel of $\sim$60 at 5500\,\AA. 

The spectrum was reduced and extracted using the standard HARPS Data Reduction Software \citep[DRS;][]{Baranne1996}. This much higher resolution spectrum displayed no obvious signs of binarity, and was used to obtain the final estimate of the stellar parameters, using the method described  Section~\ref{subsec:stellar}) below. 

%\textcolor{red}{We note that the absolute RV from both instruments (${\rm RV_{ANU}} = XX at phase XX and ${\rm RV_{HARPS}} = XX$ at phase XX) are consistent within 1\,\kms. This can also act as an independent check to rule out the eclipsing binary scenario.}

\subsection{Stellar parameters}
\label{subsec:stellar}

We used the python package \texttt{iSpec} \citep{Blanco2014} to derive the stellar effective temperature, \teff, as well as the surface gravity, \logg, and stellar metallicity, ${\rm [Fe/H]}$ from the ANU spectrum. 
We then derived the atmospheric parameters by comparing our data with synthetic spectra. Our modelling used the code \texttt{SPECTRUM} \citep{Gray1994}, with atmospheric models taken from ATLAS9 and the atomic line list from the VALD data base. \footnote{\label{vald}\url{http://vald.astro.uu.se}} \citep[][]{Ryabchikova2015}
The \texttt{iSpec} analysis produced \teff $= 5700 \pm 120$ K, $ {\rm [F/H]}= 0.05 \pm 0.10$ and \logg $= 3.85 \pm 0.04$ (gcc).

%This refers to Malcolm's analysis

We also used the Spectroscopy Made Easy \citep[\sme; ][]{Piskunov2017,Valenti1996} code to estimate the stellar parameters from the HARPS spectrum. \sme\ works by calculating the synthetic stellar spectra from grids of detailed atmosphere models and fitting them to the observations with a chi-square-minimisation approach. We used \sme\ Version 5.22 with the ATLAS12 model spectra \citep{Kurucz2013} to derive \teff , \logg,  [Fe/H] and the \vsini. All of these parameters were allowed to vary throughout the model fitting, while the micro- and macro-turbulence (\vmic\ and \vmac) were fixed through empirical calibration equations \citep{Bruntt2010,Doyle2014} valid for Sun-like stars after a first estimation of \teff. The required atomic and molecular parameters were taken from the \texttt{VALD} database. \citep[][]{Ryabchikova2015} A detailed description of the methodology can be found in \citet[][]{Fridlund2017} and \citet[][]{Persson2018}. We derived \teff\ $= 5907 \pm 150$ K, \logg\ $ = 3.86 \pm 0.14$ (gcc) from Mg I lines, \logg\ $ = 3.85 \pm 0.20$  (gcc) from Ca I lines, $[{\rm Fe/H}] = 0.10  \pm 0.10$ dex and \vsini~$= 8.2 \pm 0.9$~\kms. 

%These values, which are listed in Table~\ref{tab:star}, were adopted for all further analysis. 

The HARPS spectrum was independently analysed using the \texttt{specmatch-emp} package \citep{Yee2017}, which compares the observed spectrum with a library of $\approx 400$ synthetic spectra of FGK and M stars. This fitting routine also uses a chi-squared minimization approach to derive the stellar parameters, and yielded the values \teff $ = 6006 \pm 110$ K, $[{\rm Fe/H}] = 0.17 \pm 0.09$ dex, $R = 1.756 R_\odot$. These values are consistent with those found in the SME analysis, but the latter have slightly larger error bars, which we consider more realistic. We therefore adopted the values derived with SME, which are reported in Table~\ref{tab:star}, for all further analysis.

%End of Malcolm analysis
We derived the stellar mass, radius, and age using the on-line interface \texttt{PARAM-1.3}\footnote{\url{http://stev.oapd.inaf.it/cgi-bin/param}} with \texttt{PARSEC} stellar tracks and isochrones \citep{Bressan2012}, the \texttt{Gaia} parallax \citep[$\pi$=3.7714\,$\pm$\,0.0225 mas;][]{Brown2018}, a V-band magnitude of 10.286 \citep[][]{Munari2014} and the stellar parameters derived from the \sme\ analysis of the HARPS data. \citet[][]{Munari2014} reported an interstellar reddening consistent with zero, so we did not correct the V mag reported in Table~\ref{tab:star}.

\begin{figure}
    \centering
    \includegraphics[width=0.46\textwidth]{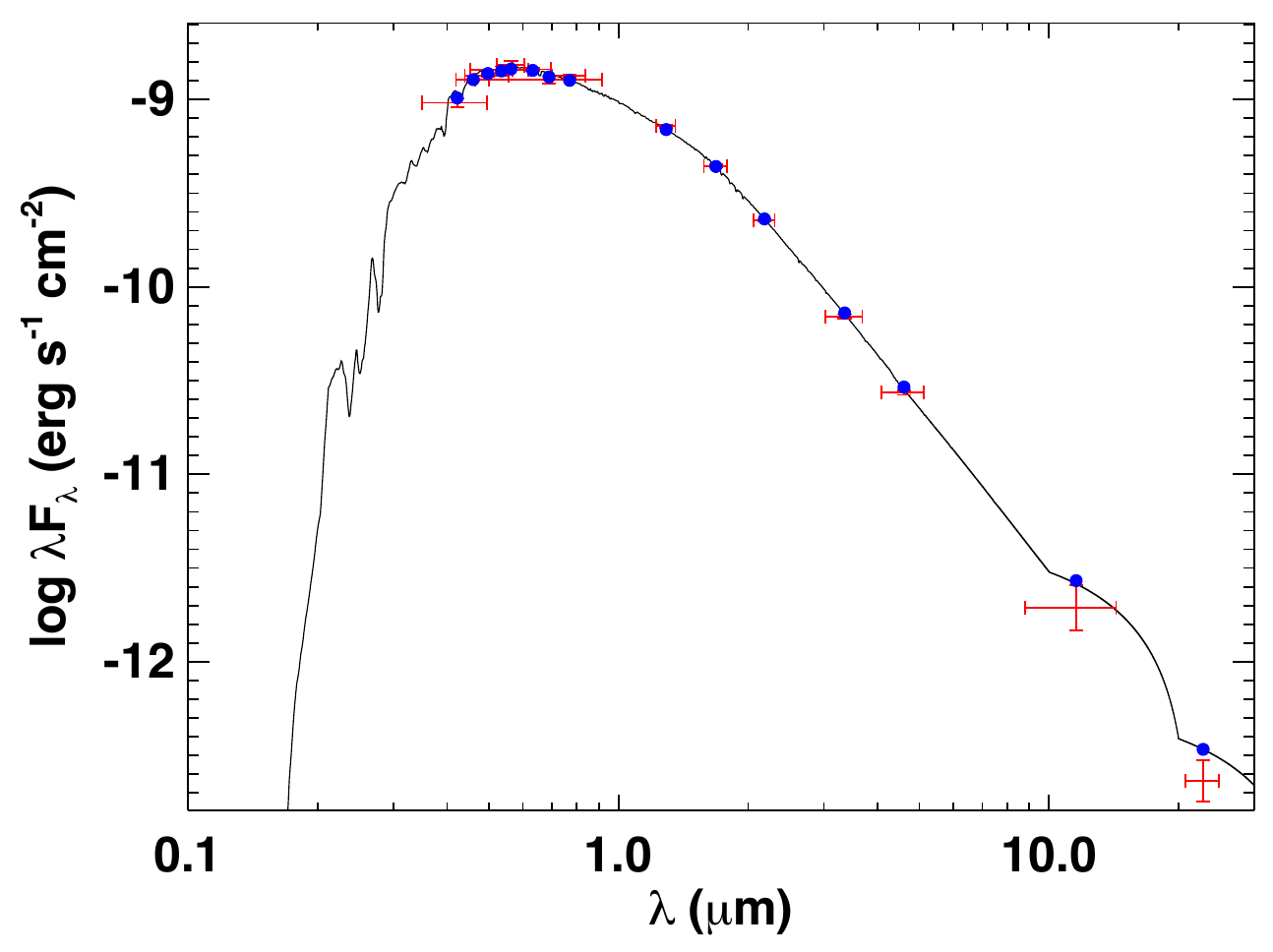} 
    \caption{Spectral energy distribution (SED) of \protect\target. Red symbols represent the observed photometric measurements, where the horizontal bars represent the effective width of the passband. Blue symbols are the model fluxes from the best-fit Kurucz atmosphere model (black).
}
    \label{fig:SED}
\end{figure}

As an independent check on the derived stellar parameters, we performed an analysis of the broadband spectral energy distribution (SED) together with the {\it Gaia\/} DR2 parallax in order to determine an empirical measurement of the stellar radius, following the procedures described in \citet{Stassun16}, \citet{Stassun17} and \citet{Stassun18}. We retrieved the $B_T V_T$ magnitudes from {\it Tycho-2}, the $BVgri$ magnitudes from APASS, the $JHK_S$ magnitudes from {\it 2MASS}, the W1--W4 magnitudes from {\it WISE}, and the $G$ magnitude from {\it Gaia} (see Table~\ref{tab:star}). Together, the available photometry spans the full stellar SED over the wavelength range 0.4--22~$\mu$m (see Figure~\ref{fig:SED}). 
We performed a fit using Kurucz stellar atmosphere models, with the priors on effective temperature ($T_{\rm eff}$), surface gravity ($\log g$), and metallicity ([Fe/H]) from the spectroscopically determined values. The remaining free parameter is the extinction ($A_V$), which we restricted to the maximum line-of-sight value from the dust maps of \citet{Schlegel98}. The resulting fits are excellent (Figure~\ref{fig:SED}) with a reduced $\chi^2$ of 2.5. The best fit extinction is $A_V = 0.00^{+0.01}_{-0.00}$. Integrating the (unreddened) model SED gives the bolometric flux at Earth of $F_{\rm bol} = 1.852 \pm 0.021 \times 10^{-9}$ erg~s~cm$^{-2}$. Taking the $F_{\rm bol}$ and $T_{\rm eff}$ together with the {\it Gaia\/} DR2 parallax, adjusted by $+0.08$~mas to account for the systematic offset reported by \citet{StassunTorres18}, gives the stellar radius as $R = 1.891 \pm 0.097$~R$_\odot$. Finally, estimating the stellar mass from the empirical relations of \citet{Torres10} gives $M = 1.39 \pm 0.10 M_\odot$, which with the radius gives the mean stellar density $\rho = 0.290 \pm 0.049 $ g~cm$^{-3}$. These agree well with the spectroscopically derived parameters.

As a subgiant, \target\ is expected to display solar-like (p-mode) oscillations which, if detected, could provide an independent estimate of the stellar parameters. We performed a search for such oscillations using the Lomb-Scarge periodogram \citep{Lomb76,Scargle82}, but did not detect any evidence of oscillations. The TESS Asteroseismology Consortium (TASC) were also unable to detect oscillations  in this object (W. Chaplin, priv. comm.).

\section{Planet validation}
\label{sec:validation}

\subsection{High Resolution Imaging} \label{subsec:highresim}

We performed speckle imaging using the Zorro instrument on the 8.1-m Gemini South telescope \citep{Matson2019} in order to search for close companions and to quantify their contribution to the \tess\ photometric aperture. Observations were obtained on the  night of 16 July 2019 using simultaneous two-color diffraction-limited optical imaging with 60 msec exposures in sets of 1000 frames. The fast exposure times and rapid read out effectively "freeze out" atmospheric turbulence. The image was reconstructed in Fourier space, a standard method for speckle image processing \citep{Howell2011}. We detect no companions within 1.17 $\arcsec$ of the target at the 4--5 $\Delta$ mag limit at 562 nm and 5--7 $\Delta$ mag limit at 832 nm. 

Additionally, we searched for nearby sources to \target\ with speckle images obtained using the HRCam speckle imager on the 4.1-m Southern Astrophysical Research \citep[SOAR; ][]{Tokovinin2018} telescope at Cerro Pachon Observatory. The \textit{I}-band observations, obtained as part of the SOAR \tess Survey \citep{Ziegler2019} on the  night of the 14 July 2019, show no evidence of any faint companions within 3 $\arcsec$ of \target\ up to a magnitude difference of 7 mag.

\begin{figure}
    \centering
    \includegraphics[width=0.45\textwidth]{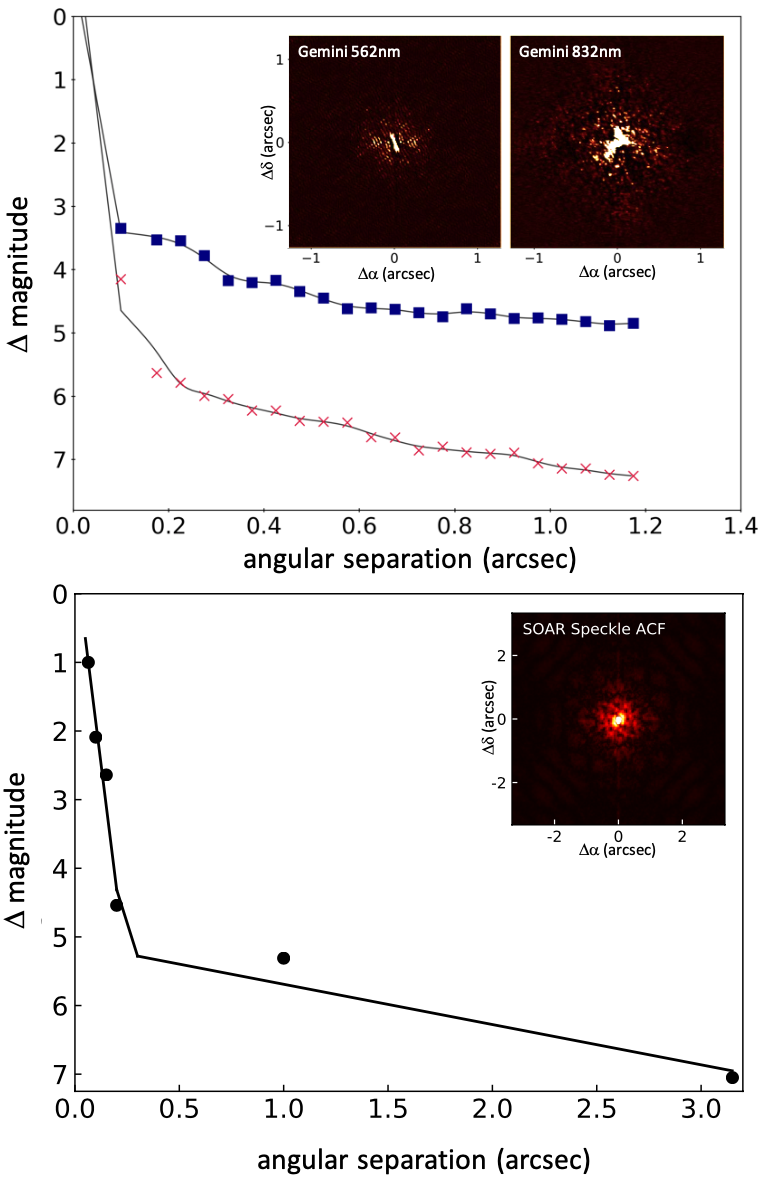}
    \caption{Contrast curves showing the 5 $\sigma$ detection sensitivity and speckle auto-correlation functions. The top panel shows the data obtained using Zorro on Gemini with the filters centered on 562 nm (blue squares) and 832 nm (red crosses) and the bottom panel shows the observations obatined with the HRCam speckle imager on SOAR in the \textit{I}-filter.
    }
    \label{fig:SOARcc}
\end{figure}

The 5 $\sigma$ detection sensitivity and the speckle auto-correlation function from the SOAR and Gemini observation, as well as the speckle auto-correlation functions, are plotted in Figure~\ref{fig:SOARcc}.

\subsection{Radial Velocity Monitoring}
\label{subsec:RV}

We collected a total of four spectra of \target\ using the CHIRON high-resolution echelle spectrograph \citep{Tokovinin2013} between 13 September and 26 October 2019, in order to confirm that the mass of \targetb\ is consistent with that of a planetary body ($M < 13 M_{\rm Jup}$). The CHIRON instrument, which is mounted on the CTIO/SMARTS 1.5-meter telescope at the Cerro Tololo Inter-American observatory in Chile, has a 2.7 arcsec diameter multi-mode optical fibre and we used the high efficient image slicer mode in order to obtain a spectral resolution of R $\sim$ 80,000. The median SNR across the four observations was found to be around 15-20 for the median order. The observations were wavelength calibrated by the CHIRON pipeline and corrected for night drift using Th-Ar spectra, which were taken before and after each science exposure. The correction for the motion of the Earth about the barycentre of the Solar system was calculated using Barycorrpy \citep{barycorrpy2018}. The 59 individual orders were normalised by fitting each one with a sine-squared blaze function plus a low-order polynomial to represent the stellar continuum, using iterative $k-$sigma clipping to exclude strong absorption lines from the fit.

We extracted relative Radial Velocity (RV) measurements from the spectra using a  novel, template free approach based on Gaussian Processes (GP) regression. Specifically, all available spectra were fit using a single GP model (using a Matern 3/2 kernel), allowing for small shifts in RV between the different  epochs. This approach makes no assumptions about the stellar spectrum, except that it is not intrinsically variable. The number of free parameters of the fit were thus $N+1$, where $N=4$ is the number of epochs: $N-1$ RV shifts and 2 GP covariance parameters (input and output scales). After performing a maximum-likelihood fit to obtain an initial estimate of the hyper-parameters, we used the \textit{emcee} Markov Chain Monte Carlo package \citep{emcee} to compute median RV estimates and their $1-\sigma$ confidence intervals (using improper uniform priors for the RV shifts, and log-uniform priors for the GP covariance parameters). The RV extraction was performed independently for each of the 59 echelle orders. At this stage we identified and excluded 30 orders which systematically gave highly discrepant RVs (more than 5 standard deviations away from the median of the other orders), before combining the remaining RV estimates into a final value for each epoch using inverse variance weighting. The resulting three relative RV measurements and their uncertainties are listed in Table~\ref{tab:RV}.

\begin{table}
\centering
    \caption{Relative radial velocity measurements of \target. \label{tab:RV}}
    
    \begin{tabular}{ll}
    \hline
    RV (\ms)    & Uncertainties (\ms) \\
    \hline
    0 & - \\
    12.6 & 27.2  \\
    21.7 & 30.6  \\
    18.7 & 27.9  
    \end{tabular}
\end{table}

\begin{figure}
    \centering
    \includegraphics[width=0.45\textwidth]{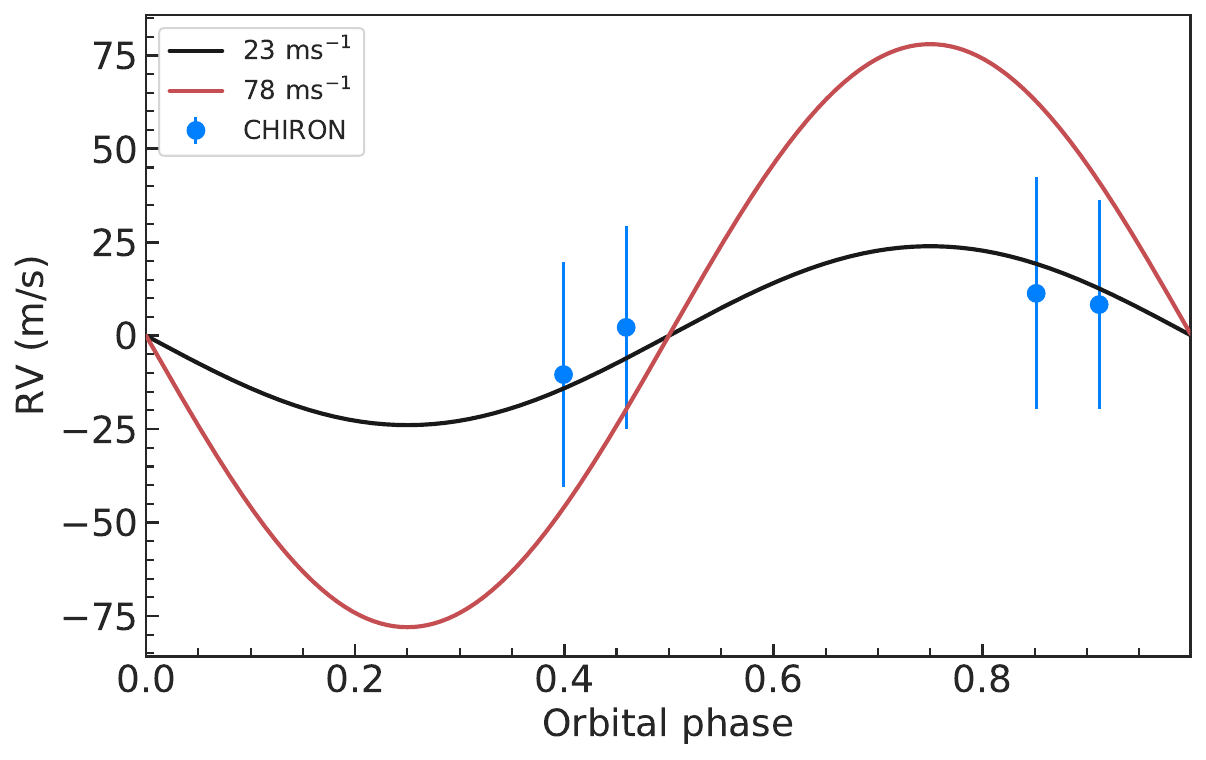}
    \caption{CHIRON RV data (blue points with error bars). A model with semi-amplitude of 23 \ms\ (corresponding to the median of he credible interval) is shown with a black line.
    A model with semi-amplitude of 78 \ms\ (corresponding to the 99\% credible interval) is shown with a red line.}
    \label{fig:rvs}
\end{figure}

The RV measurements were then used to constrain the maximum RV amplitude induced by the transiting companion. This was done by fitting for a circular orbit using the ephemeris found in the transit analysis (see Section~\ref{subsec:modelling}). The open source code \pyan\ \citep{pyaneti} was used to sample for an instrumental offset and the induced Doppler semi-amplitude, adopting a uniform prior on the semi-amplitude between 0 and 1 \kms. Figure \ref{fig:rvs} shows the CHIRON data together with the inferred model. We obtained a median value of the semi-amplitude of 23 \ms\ with a limit (at the 99 \% credible interval) of 78 \ms, corresponding  to a maximum planet mass of 2\,$M_{\rm Jup}$. This  is consistent with a planet scenario. We note that this is a preliminary analysis and that a more detailed approach would likely yield more precise RV measurements from the same data. Nonetheless, the constraints that we are able to place using this method are efficient to confirm the planetary nature of \targetb.

\subsection{Statistical validation}
\label{subsec:vespa}

The most direct way to confirm the planetary nature of \targetb\ would be to measure the radial velocity (RV) wobble it induces in its host star, and thereby measure the companion's mass. However, for such a long-period object this will take at least a full season of RV observations, and is beyond the scope of this paper. On the other hand, we are in a position to validate the planetary nature of \targetb\ in the statistical sense, by evaluating the likelihood that the transit signals are caused by a planet as opposed to the range of alternative, astrophysical false positive scenarios, using all of the available data. Statistical validation of transiting planets became routine in the era of \textit{Kepler} \citep[e.g.,][]{borucki12, morton12, diaz14, santerne15}, as many of the systems detected by that satellite are too faint for high-precision RV follow-up, and can be applied in much the same way to \tess\ targets. 

We used the open source package \vespa\ \citep{morton12, Morton2015}, which calculates the false positive probability (FPP) by considering three astrophysical false positive scenarios: undiluted eclipsing binary (EB); eclipsing binary that is diluted by another star, often known as a background eclipsing binary (BEB); and hierarchical triple eclipsing binary (HEB). We used version 0.4.7 of the software (with the MultiNest backend) with the following inputs: the 2-minute cadence detrended phase folded \tess\ light curve, the stellar parameters as derived in Section~\ref{subsec:stellar} and listed in Table~\ref{tab:star}, the two contrast curves derived from the SOAR and Gemini speckle images (Section~\ref{subsec:highresim}), and the upper limit of the depth of a potential secondary eclipse  (Section~\ref{subsec:vetting}). We found the FPP to be 0.003, which is below the commonly used validation threshold of FPP < 0.01 \citep[e.g.,][]{Livingston2018,Montet2015,Morton2016}, allowing us to conclude \targetb\ is a non-self-luminous object transiting the main target star.

\section{Results and discussion} \label{sec:resdic}

\begin{table*}
\centering
  \caption{System parameters. \label{tab:parstarget}}  
  \begin{tabular}{lcc}
  \hline
  Parameter & Prior$^{(\mathrm{a})}$ & Value$^{(\mathrm{b})}$  \\
  \hline
  \multicolumn{3}{l}{\emph{Model Parameters for  \protect\targetb}} \\
  \noalign{\smallskip}
    Orbital period $P_{\mathrm{orb}}$ (days)  &  $\mathcal{U}[  83.7945 , 83.9859]$ & \Pb[] \\
    Transit epoch $T_0$ (BJD - 2,457,000)  & $\mathcal{U}[ 1454.5781 , 1454.7696]$ & \Tzerob[]  \\
    $\sqrt{e}\sin \omega $  &  $\mathcal{F}[0]$ & 0  \\
    $\sqrt{e}\cos \omega $  &  $\mathcal{F}[0]$ & 0  \\
    Scaled semi-major axis $a/R_{\star}$ &   $\mathcal{U}[1.1,100]$ & \arb[]\\
    Scaled planet radius  $R_\mathrm{p}/R_{\star}$ &  $\mathcal{U}[0,0.06]$ & \rrb[]  \\
    Impact parameter, $b$ &  $\mathcal{U}[0,1]$  & \bb[] \\
    Parameterized limb-darkening coefficient $q_1$  & $\mathcal{U}[0,1]$ & \qoneSC  \\
    Parameterized limb-darkening coefficient $q_2$ & $\mathcal{U}[0,1]$ & \qtwoSC \\
    \hline
    \multicolumn{3}{l}{\emph{Derived parameters}} \\
  \noalign{\smallskip}
    %Planet mass ($M_{\oplus}$)  & $\cdots$ & \mpb[] \\
    Planet radius ($R_{\oplus}$)  & $\cdots$ & \rpb[] \\
    Orbit eccentricity $e$ & $\cdots$ & 0  \\
    semi-major axis $a$ (AU)  & $\cdots$ & \ab[] \\
    Orbit inclination $i$ (deg)  & $\cdots$ & \ib[] \\
    Stellar density $\rho_\star$ (from LC) & $\cdots$ & \denstrb[] \\
    Equilibrium temperature (albedo = 0) $T_{\rm eq}$ ($K$)   & $\cdots$ & \Teqb[] \\
    Insolation $F_{\rm p}$ ($F_{\oplus}$)   & $\cdots$ & \insolationb[] \\
    \hline
   \noalign{\smallskip}
  \end{tabular}
  \begin{tablenotes}\footnotesize
  \item \emph{Note} -- $^{(\mathrm{a})}$ $\mathcal{U}[a,b]$ refers to uniform priors between $a$ and $b$, $\mathcal{N}[a,b]$ to Gaussian priors with median $a$ and standard deviation $b$, and $\mathcal{F}[a]$ to a fixed value $a$.  
  $^{(\mathrm{b})}$  Inferred parameters and errors are defined as the median and 68.3\% credible interval of the posterior distribution.
\end{tablenotes}
\end{table*}

Having established that the transits are almost definitely caused by a planet, we proceed to derive the parameters of the planet by detailed modelling of the \tess\ light curve, combined with the stellar parameters derived in Section~\ref{sec:stpar}.

\subsection{Transit modelling} \label{subsec:modelling}

\begin{figure}
    \centering
    \includegraphics[width=0.45\textwidth]{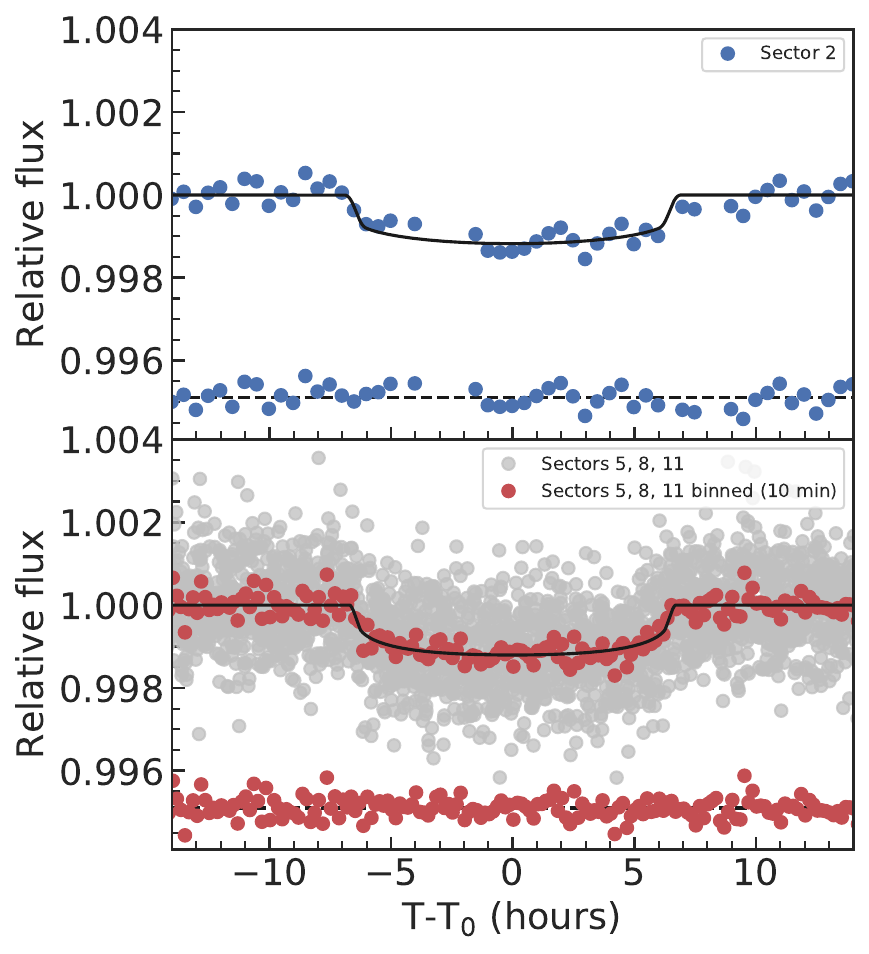}
    \caption{
    Phase-folded \tess\ light curve of \protect\targetb\ with the best-fit transit model overplotted and residuals.}
    \label{fig:transits_ff}
\end{figure}

The \target\ transits were modelled using the open source software \pyan\ \citep{pyaneti}, which was previously used for the analysis of other exoplanets discovered by \tess\ \citep[e.g.\ ][]{Gandolfi2018,Esposito2019}. We first isolated and flattened each transit using \texttt{exotrending} \citep{exotrending}  as described by \citet{Barragan2018b}. When modelling the Sector 2 transit, the variations in the transit model during each 30-min exposure must be accounted for explicitly \citep{Kipping2010}; we thus computed the model at 3-min intervals and integrated it to 30-min sampling before comparing it to the observations. Details of the fitted parameters and the priors used are given in Table~\ref{tab:parstarget}. Note that we kept the eccentricity fixed to zero for the fiducial analysis, and used the ($q_1,q_2$) limb-darkening parametrisation to sample efficiently physical solution ranges \citep[see ][]{Kipping2013}. 
%\textcolor{red}{Any particular reason?} 
The parameter space was explored using a Markov Chain Monte Carlo (MCMC) approach with 500 chains. Once the chains had converged, we used the last 5000 iterations with a thin factor of 10 to create posterior distributions based on 250,000 sampled points for each fitted parameter.

We inspected the posterior distributions visually and found them to be smooth and unimodal for all fitted parameters. We also inspected the residuals between the model and data and conclude that they show no evidence of correlated noise. We also find no evidence of enhanced scatter in the residuals.
The phase-folded data and best-fit model are shown in Figure~\ref{fig:transits_ff}, and the median and $68.3\%$ central interval of the posterior distribution for each parameter are listed in Table~\ref{tab:parstarget}. 

%\textcolor{red}{In this section we should discuss whether the residuals are Gaussian, whether they show any evidence of correlated noise, whether their scatter is comparable to the expected noise level, and whether there is any evidence of enhanced scatter in the residuals during the transit (as opposed to just outside the transits) which would arise e.g. from spot crossings.}

%!!!!!!!!!!!!!!!!!!!!!!!!!!!!!!!!!!
%\textcolor{red}{Show some of the parameter posterior distributions, perhaps in an online-only figure?} 
%!!!!!!!!!!!!!!!!!!!!!!!!!!!!!!!!!!

After fitting all four transits simultaneously, we fitted each of the four transits in turn, keeping all parameters fixed except for the time of transit centre. The resulting transit times were (in units of ${\rm BJD} - 2\,457\,000$) \Tzerobone, \Tzerobtwo, \Tzerobthree\ and \Tzerobfour. The larger uncertainty on the time of first transit is due to the longer cadence of the Sector 2 observations. We note that these values are consistent with a constant period.
%\textcolor{red}{Are those transit times consistent with a perfectly periodic ephemeris? If so, just note it, if not we would need to do additional tests to establish any evidence for TTVs.}

The stellar density derived from the transit modelling (\denstrb) is in good agreement with the value derived from the spectral analysis, (\densspb) implying that the assumption of zero eccentricity is justified, or at the very least consistent with the available data. We did however test the effect of relaxing this assumption, by repeating the transit modelling with a free eccentricity. The model was run with uniform priors over the interval $[-1,1]$ for the parameters $\sqrt{e} \sin \omega$ and $\sqrt{e} \cos \omega$ \citep{Anderson2011}, and a prior on $a/R_\star$ that was based on the stellar density derived in Section~\ref{subsec:stellar} and on Kepler's third law. This resulted in fitted values of $\sqrt{e} \sin \omega = 0.00 \pm 0.15$  and $\sqrt{e} \cos \omega = 0.04 _{-0.22}^{+0.20}$, which translate to an eccentricity of $e=0.05_{-0.03}^{+0.06}$, with an upper limit of $e<0.2$ at the 99\% confidence level.

Data from the NASA Exoplanet Archive \citep{Akeson2013} show that single planetary systems with orbital periods within 15\% of that of \targetb\ have a median eccentricity of 0.23. The apparently low eccentricity of \targetb\ is characteristic of planets in multiple transiting systems \citep{VanEylen2019}. This result, together with \targetb's long orbital period and the high fraction of main-sequence stars that host nearly coplanar multi-planet systems \citep[e.g.,][]{rowe14}, motivated us to search for additional transits in the \tess\ light curve, as described in Section~\ref{sec:injectiontest}, but none were found. However, a large fraction of planets smaller than $\sim 5\,R_\oplus$ and/or on periods longer than $\sim 60$ days would have been missed by such a search, as would any planets or orbits that were not co-planar enough with that of \targetb\ to transit. The possibility that \target\ is a multi-planet system remains open, and merits future exploration (for example using RV measurements). 

%\textcolor{red}{SHORTEN?}

\begin{figure}
    \centering
    \includegraphics[width=0.46\textwidth]{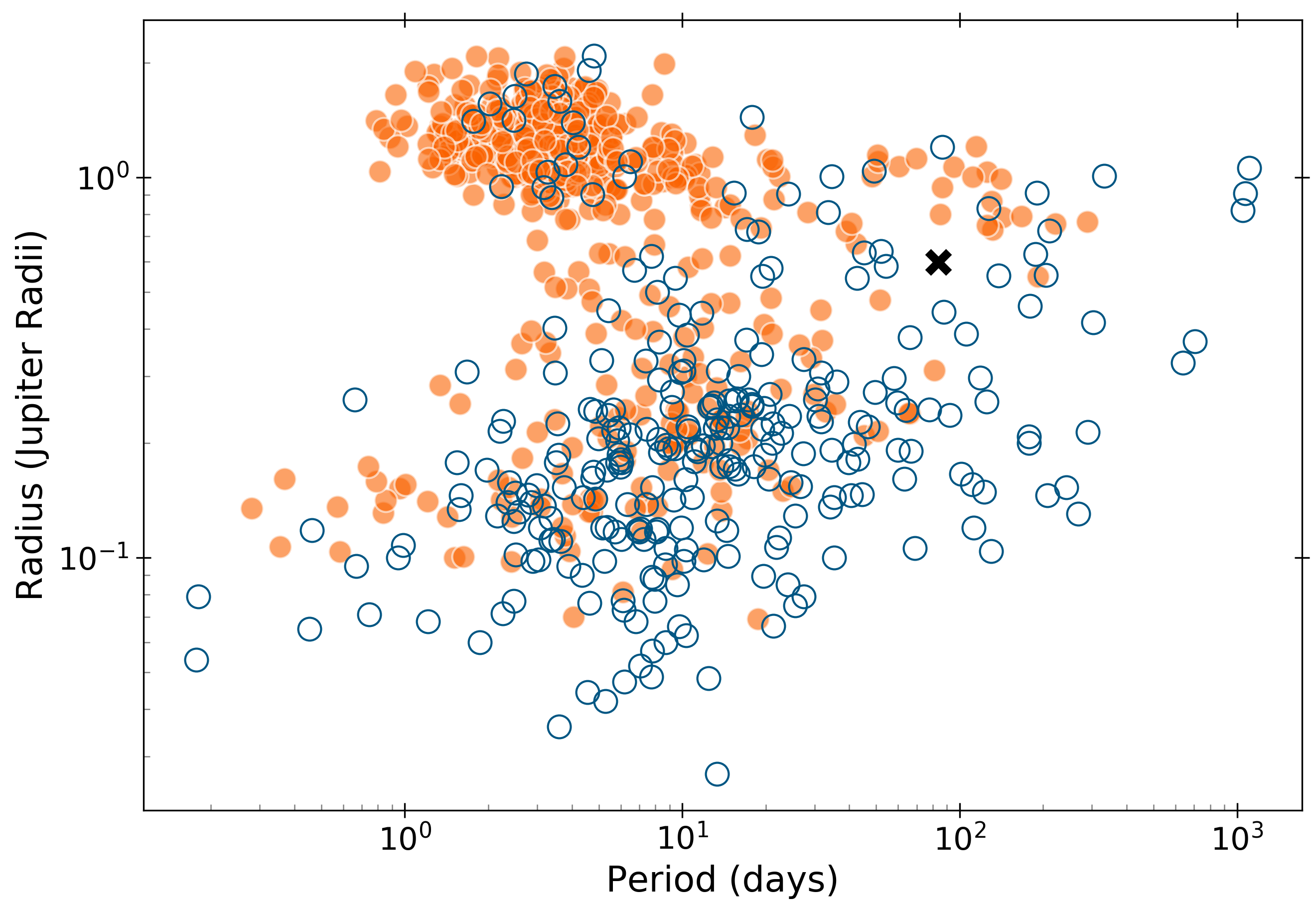}
    \caption{A log-log plot of the period vs radius of transiting exoplanets as they appear in the TEPCAT catalogue \protect\citep{tepcat}. The filled orange circles represent planets that have an at least 3-sigma mass measurement. The black cross denotes the properties of \protect\targetb.
    }
    \label{fig:rpplot}
\end{figure}

%Finish this subsection with a sentence or two about what makes this planet noteworthy --if anything-- compared to other known transiting planets, with reference to Figure~\ref{fig:rpplot}. The takeway points are: this is one of very few long-period transiting planets around bright stars, and it's around an evolved star.}
\subsection{Planets around subgiant stars} \label{subsec:subgiants}

%Planetary migration theories can be further explored by combining the results from the Rossiter-McLaughlin effect with tidal evolution models \citep[e.g.,][]{Barragan2016}, which can inform on the expansion of the stellar radius over time and the consequent fate of the planet. These models highlight the close link between the evolution of planetary systems and the evolution of their host stars.

\targetb\ is in orbit around a subgiant host. The subgiant phase in some respects mimics the pre-main sequence phase in reverse: the same star-planet interaction mechanisms -- tides, orbital migration, stellar heating and photoevaporation -- are at play, but their effect is increasing, rather than decreasing, with time. Using a MIST Version-1.2 stellar evolution track \citep{choi2016}, with input stellar parameters as given in Table~\ref{tab:star} and a rotation velocity of 0 \ms, we estimate the main sequence lifetime of \target\ to be $\sim$3.45 Gyrs. This means that, at the current age of \sage, the target has only recently left the main sequence and is still in the process of entering what will become a very rapid period of evolution. 

\targetb\ is relatively distant from its host star (scaled semi-major axis $a/R_{\star}=$ \arb) and is not expected to interact with it very strongly. Specifically, there is no reason to believe that the orbit of \targetb\ has been affected significantly by tidal interaction during the main-sequence lifetime of the star  \citep[e.g.,][]{Lanza2016}, or that the planet's size and composition have been altered in a major way by stellar irradiation \citep[e.g.,][]{Vazan2013}. However, as the star evolves further along the red giant branch the stellar luminosity and radius will increase. Using the same MIST stellar evolution track as for the analysis outlined above we estimate that the star will reach a maximum radius of 0.76 au on the red giant branch. Furthermore, the model showed that in the absence of orbital evolution of the planet, \targetb\ will be engulfed at an age of $\sim$4.66 Gyrs.

%textcolor{red}{Teq / insolation?} textcolor{red}{what about the maximum Teq / insolation?} - As it is engulfed there is no maximum?

While the radius of the star increased on the red giant branch, the mass can be assumed to remain near constant. We therefore estimate that the star's main sequence radius was $\sim 1.1 R_{\odot}$ using a mass-radius relation for main sequence stars $R \varpropto\ M ^{3/7}$ (derived from mass/radius measurements of 190 binary systems by \citet{Torres2010}. This implies that the transit of \targetb\ is already $\sim 4$ times shallower than it used to be,  highlighting the difficulties associated with detecting transiting planets around evolved host stars.

Even though the number of confirmed planets around evolved stars remains small, there is growing evidence that these systems' properties differ from those of their main sequence counterparts \citep[e.g.,][]{Johnson07,luhn19}. We explored this by evaluating the properties of planets around subgiant stars in the NASA Exoplanet Archive, where the subgiants were identified using the data driven boundaries in effective temperature and surface gravity outlined in \citet{Huber2016}.

%  \begin{equation}
%    \text{log(g)} >
%    \begin{cases}
%      13.463 - 0.00191 \text{ T}\textsubscript{eff} & \text{for}\ \text{ %T}\textsubscript{eff} > 500 K \\
%      3.9 & \text{otherwise}
%    \end{cases}
%  \end{equation}
%  
%and
%
%  \begin{equation}
%    \text{log(g)} <
%      \frac{1}{4.671} \text{ arctan} \left(\frac{\text{ %T}\textsubscript{eff} - 6300}{-67.172}\right) + 3.876 \\
%  \end{equation}

Out of the 4043 confirmed exoplanets listed in the archive, as of 2019 September, 703 have a subgiant host, 344 of which were discovered using the transit method. Out of these, only 130 have a mass measurement with an accuracy better than 3 $\sigma$. $\sim$42\% of these with measured semi-major axes lie beyond 0.5 AU of the host star, compared to $\sim$15\% for planets around dwarf stars, highlighting that detected planets around subgiants tend to have longer orbital periods. We found no significant differences in the eccentricities of the planets around subgiants compared to those around dwarfs in the sample.

Other validated planets in orbit around bright subgiant host stars observed by \tess\ include HD 1397b \citep{Brahm2019} and HD 221416 \citep{Huber2019}.

%Next, we compared the properties of \targetb to those of other planets around subgiants and found that only 25\% of the sample had orbital periods greater than 84-d. 

%\textcolor{red}{ Compare to detailed example? Look at Kepler-21.}

%The most obvious difference is the observed number of close-in planets: $<10\%$ of evolved stars host planets within $0.5\,$AU, whereas $\sim 75\%$ of planets around main sequence stars orbit within this distance of their host \citep{veras16,Bowler10}.

%\textcolor{red}{Or, alternatively, what fraction of planets around evolved stars are beyond 0.5 AU.}

%The data driven boundaries in effective temperature and surface gravity as outlined in \citet{Huber2016} were used in order to identify planets orbiting around subgiant hosts listed in the NASA Exoplanet Archive. Out of the 4043 confirmed exoplanets listed in this archive, we found 703 to have a subgiant host, 344 of which were discovered using the transit method. Out of these, 130 have a mass measurement with an accuracy better than 2 sigma. 

%Paper sent by Suzanne - low luminosity red giant branch 
%planet in an old system - probably been stable for billions of %years.
% Oscar: Check Schlichting 2014, is it useful for N1b?

\subsection{Follow-up prospects}

\targetb\ is of particular interest due to its long orbital period compared to other targets found by \tess. Accurate mass measurements of long-period transiting planets around bright stars are rare, as shown in Figure~\ref{fig:rpplot}, as are detailed studies of planets orbiting around evolved stars. Additional observations of this target can therefore help to explore a relatively under explored region of parameter space.

Future RV measurements will allow us to constrain the mass, and therefore density, of \targetb. Using \texttt{forecaster} \citep{Chen2017} we estimate the planet to have a mass of $42^{+49}_{-19} M_\oplus$ corresponding to a Doppler semi-amplitude of $5.6^{+5.3}_{-2.6}$\ms, and thus making it a good target for RV follow-up with state-of-the-art spectrographs in the southern hemisphere such as HARPS. These predictions are consistent with the upper mass limit derived in in Section~\ref{subsec:RV}.

%We estimate a typical error bar of \textcolor{red}{XXX} \ms for \textcolor{red}{XXX} min integrations for each data point using HARPS. This implies that with around \textcolor{red}{XXX} datapoints we can obtain a 3-$\sigma$ RV detection of the planet.

%\textcolor{red}{Should discuss expected precision of RV measurements (I think that this will not be limited by the precision of the instrument but by the $v\sin i$ of the star, perhaps Davide can help estimate the typical RV uncertainty per measurement and the likely number of RV measurements needed to measure the orbit.}

%We note that there are only 15 planets with periods equal to or longer than that of \targetb\ with mass measurements that have a precision better than 3$\sigma$ (see Figure~\ref{fig:rpplot}).}

Precise RV measurements during the transit may also reveal small deviations from the Keplerian fit in the RV curve \citep[the so-called Rossiter-McLaughlin, or RM, effect;][]{Rossiter24, McLaughlin24}. The RM effect can be used to estimate the project angle between host star's spin axis and the normal to the orbital plane of the planet \citep[e.g.,][]{Schneider2000}. While planetary migration through the disk should preserve, or even reduce, the primordial spin-orbit alignment, effects such as planet-planet scattering and Lidov-Kozai resonance \citep[][]{Kozai1962, Lidov1962} should result in a misalignment over time \citep[e.g.,][]{Storch2017, Deeg2009}. Measuring the RM effect can therefore help constrain the dynamical history of the system. For TOI-813b, we estimate the amplitude of the RM effect (which scales with the projected stellar equatorial rotational velocity \citep[\vsini ; ][]{Winn2010}, to be around \RMbLC\ and therefore detectable using instruments such as HARPS.  The RM effect combined with the stellar rotation period would allow us to measure the true 3D obliquity of the system. Unfortunately, we were unable to determine a rotation period for TOI 813 from the Lomb-Scargle periodogram (see Section~\ref{sec:stpar}).

%Additionally, \targetb\ is an excellent laboratory to test planetary formation and migration theories. Even though some models predict that planets found between 0.1 and 1 AU from their star formed \emph{in situ}, the location of \targetb\ can also be explained by planetary migration \citep[e.g.,][]{Kley2012}. One way to explore the migration history is to observe the Rossiter-McLaughlin effect \citep{Rossiter24, McLaughlin24}. We estimate the magnitude of the effect to be around \RMbLC\, and therefore detectable using instruments such as HARPS.

Long-period transiting planets such as \targetb\ are also potentially interesting targets for atmospheric follow up. \targetb\ has relatively low insolation (\insolationb) and equilibrium temperature (\Teqb, estimated using the Stefan-Boltzmann law and assuming an albedo of 0), and measuring its transmission spectrum would provide useful clues as to the atmospheric conditions in a relatively cool giant planet. To evaluate the feasibility of such an observations, we used the transmission spectroscopy metric (TSM) proposed by \cite{kempton18}. The TSM can be used as an approximation of the expected S/N for a 10 hour observation with JWST/NRIRISS based on the assumption of a cloud-free atmosphere. To calculate it for \targetb, we assumed a mass of 42 $M_\oplus$ and a mean molecular weight of 2.3. The resulting TSM of $\sim$20 is relatively low, as expected (cooler atmospheres are less puffy), but this is to some extent compensated by the relatively long duration of the transit.

%!!!!!!!!!!!!!!!!!!!
%\textcolor{red}{How does this compare with other transiting planets with similar sizes and temperatures? Is it one of the better warm Saturns to go after, or just average?}

\subsection{Prospects for PHT}

%\textbf{Shall we discuss possible different composition based on different masses?}

This is the first validated \tess\ exoplanet found by citizen scientists taking part in the PHT project, but many more possible discoveries are actively being followed up. PHT discovers planet candidates through two distinct routes: about 15\% are brought to the attention of the science team via the \textit{Talk} discussion boards, while the rest are identified by the main PHT \tess\ pipeline, which combines the classifications of multiple volunteers for each light curve, using a density based clustering algorithm (Eisner et al. in prep). Based on our preliminary findings for the first 9 \tess\ observation Sectors, we find about 5 high quality candidates per sector, which pass all the light curve based vetting tests discussed in Section~\ref{subsec:vetting} but were not found by the SPOC and QLP pipelines. Extrapolating to the two-year nominal \tess\ missions, we thus expect PHT to find over 100 new candidates over all. 

About two thirds of our candidates so far are long-period, exhibiting only a single-transit event in the \tess\ data. These are much more challenging to validate statistically, and are particularly challenging to follow up, so the few high-latitude systems that were observed by \tess\ long enough to display multiple transits are particularly valuable. Interestingly, several of our other early candidates are listed in the TIC as subgiant stars. These all have long periods and hence long durations, which may explain why more of them were missed by the standard pipelines. If this trend is confirmed, the detection of planets around evolved stars, particularly those where mass measurements are feasible, will be one of the lasting contributions of the PHT project.

%OSCAR comments:
%Only 35 exoplanet with period > 30 days have mass measurements better than 50%
%only 13 exoplanet with period > 89 days have mass measurements better than 50%
%Check if those 13 have masses derived by RVs or TTVs

%\textbf{Long period planets are import: }
%
%- Help understand migration of planets. 
%- \textit{likely formed in situ }
%- Insulation? (from Pyanti) 

%- atmospheric scale height can be estimated- done-ish.

%- Check whether we would expect it to be tidally locked - as star grows it will become tidally locked. Spin %state of the planet would be expected to change a lot. 

%- discussion about what we expect the RV signal for different compositions (mass radius diagram, lines for 
%different compositions (see if there is anything that allows us to vary the core mass. 

%vsini 6-8 km (limits precision of RV measurments)

\section{Conclusions} 
\label{sec:conclude}

We report on the discovery and validation of the first \tess\ planet that was found through a citizen science campaign. The signal was initially discovered by Planet Hunters TESS volunteers as a single transit event in the Sector 5 target pixel light curve (2-minute cadence). Three additional transits were later found in the target pixel files of sectors 8 and 11 and in the full frame images (30-minute cadence) of Sector 2. 

The candidate passed all of our light curve based vetting checks, including, but not limited to, odd-even transit depth comparison, checks for systematic effects, searches for secondary eclipses and pixel-level centroid analysis to search for blends. Further false positive scenarios, including blended eclipsing binaries, were ruled out with the aid of speckle imaging. These observations, obtained with Gemini and SOAR, showed no signs of stellar companions down to a magnitude difference of 4-5 within 1.17 \arcsec\ and down to a magnitude difference of 7 within 3 \arcsec\ of the target. Additionally, we obtained reconnaissance and high resolution spectra of the star in order to refine the stellar parameter and allowing us to statistically validate the planet, with a false positive probability of 0.003, using \vespa. Finally we determined an upper mass limit of the candidate using RV observations obtained with the CHIRON spectrograph. The upper limit of $2\,M_{\rm Jup}$ (at the 99 \% credible interval) is consistent with a planetary companion scenario.

A BLS search, carried out on the detrended light curve, did not reveal any additional transit signals. We therefore carried out injection and recovery tests to quantify the detectability of potential further planets in the system. The results showed that we were able to recover >80\% of the injected planets larger than $\sim 5 \,R_{\oplus}$ with periods $<30$\, and $\sim$50\% of the simulated planets  with periods $<60$ days, meaning that we are not able to rule out the presence of additional planets inside the orbit of \targetb. 

Detailed modelling of the transits yield that the planet has an orbital period of \Pb, a planet radius of \rpb\, and a semi major axis of \ab. Furthermore, the planet is in orbit around a bright (V = 10.3 mag) subgiant ($R_\star=1.94\,R_\odot$, $M_\star=1.32\,M_\odot$) star which is in the process of evolving away from the main sequence and onto the red giant branch. Stellar evolutionary tracks showed that the expanding stellar radius will reach the current semi-major axis of \targetb\ in $\sim$0.93 Gyrs.

The planet's relatively long orbital period together with the evolved nature of the host star places \targetb\ in a relatively under explored region of parameter space and is therefore an exciting target for follow-up observations. Based on the stellar brightness (V = 10.3 mag) and expected planetary mass ($42^{+49}_{-19} M_\oplus$), we estimate that \targetb\ induces a reflex motion with a Doppler semi-amplitude of $\sim6$\ms, making this a promising candidate for which we can obtain a precise mass measurement. 

Over the two-year \tess\ mission we expect the Planet Hunters TESS project to find over 100 new planet candidates in the 2-minute cadence light curves alone. We anticipate that some of these long-period planet candidates may be re-detected in the TESS extended mission, yielding precise orbital period measurements and paving the way for more detailed studies.

%To date, there are only 15 planets with reliable mass measurements (> 3$\sigma$) with orbital periods greater than the \Pb\ period of \targetb. 

\section*{Acknowledgements}

We like to thank all of the volunteers who participated in the Planet Hunters TESS project, as without them this work would not have been possible. 

N.L.E. thanks the LSSTC Data Science Fellowship Program, which is funded by LSSTC, NSF Cybertraining Grant number 1829740, the Brinson Foundation, and the Moore Foundation; her participation in the program has benefited this work. N.L.E. also acknowledges funding from the Science and Technology Funding Council (STFC) Grant Code
ST/R505006/1. A.V.'s work was performed under contract with the California Institute of Technology/Jet Propulsion Laboratory funded by NASA through the Sagan Fellowship Program executed by the NASA Exoplanet Science Institute. M.H.K. thanks Allan R. Schmitt and Troy Winarski for making their light curve examining softwares LcTools and AKO-TPF freely available. M.E.S. was supported by Gemini Observatory and also in part by Queen's University Belfast. A.Z. acknowledges support from an Australian Government Research Training Program (RTP) Scholarship. O.B. and S.A. acknowledge support from the UK Science and Technology Facilities Council (STFC) under grants ST/S000488/1 and ST/R004846/1. We would also like to thank the Zorro team and for their help with obtaining speckle images that allowed us to validate this planet and Andy Stephens for his insight into Zorro/'Alopeke. L. M. F. T. would like to thank the CONACyT for its support through the grant CVU 555458.

This paper includes data collected with the TESS mission, obtained from the MAST data archive at the Space Telescope Science Institute (STScI). Funding for the TESS mission is provided by the NASA Explorer Program. STScI is operated by the Association of Universities for Research in Astronomy, Inc., under NASA contract NAS 5-26555.

This research has made use of the NASA Exoplanet Archive and the Exoplanet Follow-up Observation Program website, which are operated by the California Institute of Technology, under contract with the National Aeronautics and Space Administration under the Exoplanet Exploration Program.

Resources supporting this work were provided by the NASA High-End Computing (HEC) Program through the NASA Advanced Supercomputing (NAS) Division at Ames Research Center for the production of the SPOC data products.

This work is based in part on observations from Director's Discretionary program GS-2019A-DD-109 at the Gemini Observatory, which is operated by the Association of Universities for Research in Astronomy, Inc., under a cooperative agreement with the NSF on behalf of the Gemini partnership: the National Science Foundation (United States), National Research Council (Canada), CONICYT (Chile), Ministerio de Ciencia, Tecnolog\'{i}a e Innovaci\'{o}n Productiva (Argentina), Minist\'{e}rio da Ci\^{e}ncia, Tecnologia e Inova\c{c}\~{a}o (Brazil), and Korea Astronomy and Space Science Institute (Republic of Korea).

This paper is based on observations obtained at the Southern Astrophysical Research (SOAR) telescope, which is a joint project of the Minist\'{e}rio da Ci\^{e}ncia, Tecnologia, Inova\c{c}\~{o}es e Comunica\c{c}\~{o}es (MCTIC) do Brasil, the U.S. National Optical Astronomy Observatory (NOAO), the University of North Carolina at Chapel Hill (UNC), and Michigan State University (MSU).

N.L.E. and O.B. wish to thank the charming local caf\'e in Oxford, where TOI-813b was first identified as a high-priority PHT candidate and affectionately dubbed Nora-1b, and where most of this paper was written.

%Some of the observations in the paper made use of the High-Resolution Imaging instrument Zorro at Gemini-South. Zorro was funded by the NASA Exoplanet Exploration Program and built at the NASA Ames Research Center by Steve B. Howell, Nic Scott, Elliott P. Horch, and Emmett Quigley.

%%%%%%%%%%%%%%%%%%%%%%%%%%%%%%%%%%%%%%%%%%%%%%%%%%



\bibliographystyle{mnras}
\bibliography{bibs} % if your bibtex file is called example.bib

\begin{thebibliography}{}
\makeatletter
\relax
\def\mn@urlcharsother{\let\do\@makeother \do\$\do\&\do\#\do\^\do\_\do\%\do\~}
\def\mn@doi{\begingroup\mn@urlcharsother \@ifnextchar [ {\mn@doi@}
  {\mn@doi@[]}}
\def\mn@doi@[#1]#2{\def\@tempa{#1}\ifx\@tempa\@empty \href
  {http://dx.doi.org/#2} {doi:#2}\else \href {http://dx.doi.org/#2} {#1}\fi
  \endgroup}
\def\mn@eprint#1#2{\mn@eprint@#1:#2::\@nil}
\def\mn@eprint@arXiv#1{\href {http://arxiv.org/abs/#1} {{\tt arXiv:#1}}}
\def\mn@eprint@dblp#1{\href {http://dblp.uni-trier.de/rec/bibtex/#1.xml}
  {dblp:#1}}
\def\mn@eprint@#1:#2:#3:#4\@nil{\def\@tempa {#1}\def\@tempb {#2}\def\@tempc
  {#3}\ifx \@tempc \@empty \let \@tempc \@tempb \let \@tempb \@tempa \fi \ifx
  \@tempb \@empty \def\@tempb {arXiv}\fi \@ifundefined
  {mn@eprint@\@tempb}{\@tempb:\@tempc}{\expandafter \expandafter \csname
  mn@eprint@\@tempb\endcsname \expandafter{\@tempc}}}

\bibitem[\protect\citeauthoryear{{Aigrain} \& {Irwin}}{{Aigrain} \&
  {Irwin}}{2004}]{Aigrain04}
{Aigrain} S.,  {Irwin} M.,  2004, \mn@doi [\mnras]
  {10.1111/j.1365-2966.2004.07657.x}, \href
  {https://ui.adsabs.harvard.edu/abs/2004MNRAS.350..331A} {350, 331}

\bibitem[\protect\citeauthoryear{{Akeson} et~al.,}{{Akeson}
  et~al.}{2013}]{Akeson2013}
{Akeson} R.~L.,  et~al., 2013, \mn@doi [\pasp] {10.1086/672273}, \href
  {https://ui.adsabs.harvard.edu/abs/2013PASP..125..989A} {125, 989}

\bibitem[\protect\citeauthoryear{{Anderson} et~al.,}{{Anderson}
  et~al.}{2011}]{Anderson2011}
{Anderson} D.~R.,  et~al., 2011, \mn@doi [\apjl] {10.1088/2041-8205/726/2/L19},
  \href {https://ui.adsabs.harvard.edu/abs/2011ApJ...726L..19A} {726, L19}

\bibitem[\protect\citeauthoryear{{Baranne} et~al.,}{{Baranne}
  et~al.}{1996}]{Baranne1996}
{Baranne} A.,  et~al., 1996, \aaps, \href
  {https://ui.adsabs.harvard.edu/abs/1996A&AS..119..373B} {119, 373}

\bibitem[\protect\citeauthoryear{{Barclay}, {Pepper}  \& {Quintana}}{{Barclay}
  et~al.}{2018}]{Barclay2018}
{Barclay} T.,  {Pepper} J.,   {Quintana} E.~V.,  2018, \mn@doi [\apjs]
  {10.3847/1538-4365/aae3e9}, \href
  {https://ui.adsabs.harvard.edu/abs/2018ApJS..239....2B} {239, 2}

\bibitem[\protect\citeauthoryear{{Barrag{\'a}n} \& {Gandolfi}}{{Barrag{\'a}n}
  \& {Gandolfi}}{2017}]{exotrending}
{Barrag{\'a}n} O.,  {Gandolfi} D.,  2017, {Exotrending} (\mn@eprint {ascl}
  {1706.001})

\bibitem[\protect\citeauthoryear{{Barrag{\'a}n} et~al.,}{{Barrag{\'a}n}
  et~al.}{2018}]{Barragan2018b}
{Barrag{\'a}n} O.,  et~al., 2018, \mn@doi [\aap] {10.1051/0004-6361/201732217},
  \href {https://ui.adsabs.harvard.edu/abs/2018A&A...612A..95B} {612, A95}

\bibitem[\protect\citeauthoryear{Barrag\'an, Gandolfi  \&
  Antoniciello}{Barrag\'an et~al.}{2019}]{pyaneti}
Barrag\'an O.,  Gandolfi D.,   Antoniciello G.,  2019, \mn@doi [\mnras]
  {10.1093/mnras/sty2472}, \href
  {https://ui.adsabs.harvard.edu/#abs/2019MNRAS.482.1017B} {482, 1017}

\bibitem[\protect\citeauthoryear{{Blanco-Cuaresma}, {Soubiran}, {Heiter}  \&
  {Jofr{\'e}}}{{Blanco-Cuaresma} et~al.}{2014}]{Blanco2014}
{Blanco-Cuaresma} S.,  {Soubiran} C.,  {Heiter} U.,   {Jofr{\'e}} P.,  2014,
  \mn@doi [\aap] {10.1051/0004-6361/201423945}, \href
  {https://ui.adsabs.harvard.edu/abs/2014A&A...569A.111B} {569, A111}

\bibitem[\protect\citeauthoryear{{Borucki} et~al.,}{{Borucki}
  et~al.}{2012}]{borucki12}
{Borucki} W.~J.,  et~al., 2012, \mn@doi [\apj] {10.1088/0004-637X/745/2/120},
  \href {https://ui.adsabs.harvard.edu/abs/2012ApJ...745..120B} {745, 120}

\bibitem[\protect\citeauthoryear{{Boyajian} et~al.,}{{Boyajian}
  et~al.}{2016}]{boyajian16}
{Boyajian} T.~S.,  et~al., 2016, \mn@doi [\mnras] {10.1093/mnras/stw218}, \href
  {https://ui.adsabs.harvard.edu/abs/2016MNRAS.457.3988B} {457, 3988}

\bibitem[\protect\citeauthoryear{{Brahm} et~al.,}{{Brahm}
  et~al.}{2019}]{Brahm2019}
{Brahm} R.,  et~al., 2019, \mn@doi [\aj] {10.3847/1538-3881/ab279a}, \href
  {https://ui.adsabs.harvard.edu/abs/2019AJ....158...45B} {158, 45}

\bibitem[\protect\citeauthoryear{{Bressan}, {Marigo}, {Girardi}, {Salasnich},
  {Dal Cero}, {Rubele}  \& {Nanni}}{{Bressan} et~al.}{2012}]{Bressan2012}
{Bressan} A.,  {Marigo} P.,  {Girardi} L.,  {Salasnich} B.,  {Dal Cero} C.,
  {Rubele} S.,   {Nanni} A.,  2012, \mn@doi [\mnras]
  {10.1111/j.1365-2966.2012.21948.x}, \href
  {https://ui.adsabs.harvard.edu/abs/2012MNRAS.427..127B} {427, 127}

\bibitem[\protect\citeauthoryear{{Bruntt} et~al.,}{{Bruntt}
  et~al.}{2010}]{Bruntt2010}
{Bruntt} H.,  et~al., 2010, \mn@doi [\mnras]
  {10.1111/j.1365-2966.2010.16575.x}, \href
  {https://ui.adsabs.harvard.edu/abs/2010MNRAS.405.1907B} {405, 1907}

\bibitem[\protect\citeauthoryear{{Chen} \& {Kipping}}{{Chen} \&
  {Kipping}}{2017}]{Chen2017}
{Chen} J.,  {Kipping} D.,  2017, \mn@doi [\apj] {10.3847/1538-4357/834/1/17},
  \href {https://ui.adsabs.harvard.edu/abs/2017ApJ...834...17C} {834, 17}

\bibitem[\protect\citeauthoryear{{Childress}, {Vogt}, {Nielsen}  \&
  {Sharp}}{{Childress} et~al.}{2014}]{Childress2014}
{Childress} M.~J.,  {Vogt} F. P.~A.,  {Nielsen} J.,   {Sharp} R.~G.,  2014,
  \mn@doi [\apss] {10.1007/s10509-013-1682-0}, \href
  {https://ui.adsabs.harvard.edu/abs/2014Ap&SS.349..617C} {349, 617}

\bibitem[\protect\citeauthoryear{{Choi}, {Dotter}, {Conroy}, {Cantiello},
  {Paxton}  \& {Johnson}}{{Choi} et~al.}{2016}]{choi2016}
{Choi} J.,  {Dotter} A.,  {Conroy} C.,  {Cantiello} M.,  {Paxton} B.,
  {Johnson} B.~D.,  2016, \mn@doi [\apj] {10.3847/0004-637X/823/2/102}, \href
  {https://ui.adsabs.harvard.edu/abs/2016ApJ...823..102C} {823, 102}

\bibitem[\protect\citeauthoryear{{Christiansen} et~al.,}{{Christiansen}
  et~al.}{2018}]{Christiansen2018}
{Christiansen} J.~L.,  et~al., 2018, \mn@doi [\aj] {10.3847/1538-3881/aa9be0},
  \href {https://ui.adsabs.harvard.edu/abs/2018AJ....155...57C} {155, 57}

\bibitem[\protect\citeauthoryear{{Cutri} \& {et al.}}{{Cutri} \& {et
  al.}}{2013}]{Cutri2013}
{Cutri} R.~M.,  {et al.} 2013, VizieR Online Data Catalog, \href
  {https://ui.adsabs.harvard.edu/abs/2013yCat.2328....0C} {p. II/328}

\bibitem[\protect\citeauthoryear{{Cutri} et~al.,}{{Cutri}
  et~al.}{2003}]{2MASS2003}
{Cutri} R.~M.,  et~al., 2003, VizieR Online Data Catalog, \href
  {https://ui.adsabs.harvard.edu/abs/2003yCat.2246....0C} {p. II/246}

\bibitem[\protect\citeauthoryear{{Deeg} et~al.,}{{Deeg}
  et~al.}{2009}]{Deeg2009}
{Deeg} H.~J.,  et~al., 2009, \mn@doi [Astronomy and Astrophysics]
  {10.1051/0004-6361/200912011}, \href
  {https://ui.adsabs.harvard.edu/abs/2009A&A...506..343D} {506, 343}

\bibitem[\protect\citeauthoryear{{D{\'\i}az}, {Almenara}, {Santerne}, {Moutou},
  {Lethuillier}  \& {Deleuil}}{{D{\'\i}az} et~al.}{2014}]{diaz14}
{D{\'\i}az} R.~F.,  {Almenara} J.~M.,  {Santerne} A.,  {Moutou} C.,
  {Lethuillier} A.,   {Deleuil} M.,  2014, \mn@doi [\mnras]
  {10.1093/mnras/stu601}, \href
  {https://ui.adsabs.harvard.edu/abs/2014MNRAS.441..983D} {441, 983}

\bibitem[\protect\citeauthoryear{{Dopita}, {Hart}, {McGregor}, {Oates},
  {Bloxham}  \& {Jones}}{{Dopita} et~al.}{2007}]{Dopita2007}
{Dopita} M.,  {Hart} J.,  {McGregor} P.,  {Oates} P.,  {Bloxham} G.,   {Jones}
  D.,  2007, \mn@doi [\apss] {10.1007/s10509-007-9510-z}, \href
  {https://ui.adsabs.harvard.edu/abs/2007Ap&SS.310..255D} {310, 255}

\bibitem[\protect\citeauthoryear{{Doyle}, {Davies}, {Smalley}, {Chaplin}  \&
  {Elsworth}}{{Doyle} et~al.}{2014}]{Doyle2014}
{Doyle} A.~P.,  {Davies} G.~R.,  {Smalley} B.,  {Chaplin} W.~J.,   {Elsworth}
  Y.,  2014, \mn@doi [\mnras] {10.1093/mnras/stu1692}, \href
  {https://ui.adsabs.harvard.edu/abs/2014MNRAS.444.3592D} {444, 3592}

\bibitem[\protect\citeauthoryear{{Esposito} et~al.,}{{Esposito}
  et~al.}{2019}]{Esposito2019}
{Esposito} M.,  et~al., 2019, \mn@doi [\aap] {10.1051/0004-6361/201834853},
  \href {https://ui.adsabs.harvard.edu/abs/2019A&A...623A.165E} {623, A165}

\bibitem[\protect\citeauthoryear{{Fausnaugh}, {Huang}, {Glidden}, {Guerrero}
  \& {TESS Science Office}}{{Fausnaugh} et~al.}{2018}]{Fausnaugh18}
{Fausnaugh} M.,  {Huang} X.,  {Glidden} A.,  {Guerrero} N.,   {TESS Science
  Office} 2018, in American Astronomical Society Meeting Abstracts \#231. p.
  439.09

\bibitem[\protect\citeauthoryear{{Feinstein} et~al.,}{{Feinstein}
  et~al.}{2019}]{Feinstein2019}
{Feinstein} A.~D.,  et~al., 2019, arXiv e-prints, \href
  {https://ui.adsabs.harvard.edu/abs/2019arXiv190309152F} {p. arXiv:1903.09152}

\bibitem[\protect\citeauthoryear{{Fischer} et~al.,}{{Fischer}
  et~al.}{2012}]{fischer12}
{Fischer} D.~A.,  et~al., 2012, \mn@doi [\mnras]
  {10.1111/j.1365-2966.2011.19932.x}, \href
  {https://ui.adsabs.harvard.edu/abs/2012MNRAS.419.2900F} {419, 2900}

\bibitem[\protect\citeauthoryear{{Foreman-Mackey}, {Hogg}, {Lang}  \&
  {Goodman}}{{Foreman-Mackey} et~al.}{2013}]{emcee}
{Foreman-Mackey} D.,  {Hogg} D.~W.,  {Lang} D.,   {Goodman} J.,  2013, \mn@doi
  [\pasp] {10.1086/670067}, \href
  {https://ui.adsabs.harvard.edu/abs/2013PASP..125..306F} {125, 306}

\bibitem[\protect\citeauthoryear{{Foreman-Mackey}, {Morton}, {Hogg}, {Agol}  \&
  {Sch{\"o}lkopf}}{{Foreman-Mackey} et~al.}{2016}]{Foreman-Mackey2016}
{Foreman-Mackey} D.,  {Morton} T.~D.,  {Hogg} D.~W.,  {Agol} E.,
  {Sch{\"o}lkopf} B.,  2016, \mn@doi [\aj] {10.3847/0004-6256/152/6/206}, \href
  {https://ui.adsabs.harvard.edu/abs/2016AJ....152..206F} {152, 206}

\bibitem[\protect\citeauthoryear{{Fridlund} et~al.,}{{Fridlund}
  et~al.}{2017}]{Fridlund2017}
{Fridlund} M.,  et~al., 2017, \mn@doi [\aap] {10.1051/0004-6361/201730822},
  \href {https://ui.adsabs.harvard.edu/abs/2017A&A...604A..16F} {604, A16}

\bibitem[\protect\citeauthoryear{{Gaia Collaboration}}{{Gaia
  Collaboration}}{2018}]{gaiadr2}
{Gaia Collaboration} 2018, VizieR Online Data Catalog, \href
  {https://ui.adsabs.harvard.edu/abs/2018yCat.1345....0G} {p. I/345}

\bibitem[\protect\citeauthoryear{{Gaia Collaboration} et~al.,}{{Gaia
  Collaboration} et~al.}{2018a}]{Gaia2018}
{Gaia Collaboration} et~al., 2018a, \mn@doi [\aap]
  {10.1051/0004-6361/201833051}, \href
  {https://ui.adsabs.harvard.edu/abs/2018A&A...616A...1G} {616, A1}

\bibitem[\protect\citeauthoryear{{Gaia Collaboration} et~al.,}{{Gaia
  Collaboration} et~al.}{2018b}]{Brown2018}
{Gaia Collaboration} et~al., 2018b, \mn@doi [\aap]
  {10.1051/0004-6361/201833051}, \href
  {https://ui.adsabs.harvard.edu/abs/2018A&A...616A...1G} {616, A1}

\bibitem[\protect\citeauthoryear{{Gandolfi} et~al.,}{{Gandolfi}
  et~al.}{2018}]{Gandolfi2018}
{Gandolfi} D.,  et~al., 2018, \mn@doi [\aap] {10.1051/0004-6361/201834289},
  \href {https://ui.adsabs.harvard.edu/abs/2018A&A...619L..10G} {619, L10}

\bibitem[\protect\citeauthoryear{{Gray} \& {Corbally}}{{Gray} \&
  {Corbally}}{1994}]{Gray1994}
{Gray} R.~O.,  {Corbally} C.~J.,  1994, \mn@doi [\aj] {10.1086/116893}, \href
  {https://ui.adsabs.harvard.edu/abs/1994AJ....107..742G} {107, 742}

\bibitem[\protect\citeauthoryear{{H{\o}g} et~al.,}{{H{\o}g}
  et~al.}{2000}]{Hog2000}
{H{\o}g} E.,  et~al., 2000, Astronomy and Astrophysics, \href
  {https://ui.adsabs.harvard.edu/abs/2000A&A...355L..27H} {355, L27}

\bibitem[\protect\citeauthoryear{{Howell}, {Everett}, {Sherry}, {Horch}  \&
  {Ciardi}}{{Howell} et~al.}{2011}]{Howell2011}
{Howell} S.~B.,  {Everett} M.~E.,  {Sherry} W.,  {Horch} E.,   {Ciardi} D.~R.,
  2011, \mn@doi [\aj] {10.1088/0004-6256/142/1/19}, \href
  {https://ui.adsabs.harvard.edu/abs/2011AJ....142...19H} {142, 19}

\bibitem[\protect\citeauthoryear{{Huang} et~al.,}{{Huang}
  et~al.}{2018}]{Huang2018}
{Huang} C.~X.,  et~al., 2018, \mn@doi [\apjl] {10.3847/2041-8213/aaef91}, \href
  {https://ui.adsabs.harvard.edu/abs/2018ApJ...868L..39H} {868, L39}

\bibitem[\protect\citeauthoryear{{Huber} et~al.,}{{Huber}
  et~al.}{2016}]{Huber2016}
{Huber} D.,  et~al., 2016, \mn@doi [\apjs] {10.3847/0067-0049/224/1/2}, \href
  {https://ui.adsabs.harvard.edu/abs/2016ApJS..224....2H} {224, 2}

\bibitem[\protect\citeauthoryear{{Huber} et~al.,}{{Huber}
  et~al.}{2019}]{Huber2019}
{Huber} D.,  et~al., 2019, \mn@doi [\aj] {10.3847/1538-3881/ab1488}, \href
  {https://ui.adsabs.harvard.edu/abs/2019AJ....157..245H} {157, 245}

\bibitem[\protect\citeauthoryear{{Jenkins} et~al.,}{{Jenkins}
  et~al.}{2016}]{jenkins16}
{Jenkins} J.~M.,  et~al., 2016, in \procspie. p. 99133E,
  \mn@doi{10.1117/12.2233418}

\bibitem[\protect\citeauthoryear{{Johnson} et~al.,}{{Johnson}
  et~al.}{2007}]{Johnson07}
{Johnson} J.~A.,  et~al., 2007, \mn@doi [\apj] {10.1086/519677}, \href
  {https://ui.adsabs.harvard.edu/abs/2007ApJ...665..785J} {665, 785}

\bibitem[\protect\citeauthoryear{{Kanodia} \& {Wright}}{{Kanodia} \&
  {Wright}}{2018}]{barycorrpy2018}
{Kanodia} S.,  {Wright} J.,  2018, \mn@doi [Research Notes of the American
  Astronomical Society] {10.3847/2515-5172/aaa4b7}, \href
  {https://ui.adsabs.harvard.edu/abs/2018RNAAS...2....4K} {2, 4}

\bibitem[\protect\citeauthoryear{{Kempton} et~al.,}{{Kempton}
  et~al.}{2018}]{kempton18}
{Kempton} E. M.~R.,  et~al., 2018, \mn@doi [\pasp] {10.1088/1538-3873/aadf6f},
  \href {https://ui.adsabs.harvard.edu/abs/2018PASP..130k4401K} {130, 114401}

\bibitem[\protect\citeauthoryear{{Kipping}}{{Kipping}}{2010}]{Kipping2010}
{Kipping} D.~M.,  2010, \mn@doi [\mnras] {10.1111/j.1365-2966.2010.17242.x},
  \href {https://ui.adsabs.harvard.edu/abs/2010MNRAS.408.1758K} {408, 1758}

\bibitem[\protect\citeauthoryear{{Kipping}}{{Kipping}}{2013}]{Kipping2013}
{Kipping} D.~M.,  2013, \mn@doi [\mnras] {10.1093/mnras/stt1435}, \href
  {https://ui.adsabs.harvard.edu/abs/2013MNRAS.435.2152K} {435, 2152}

\bibitem[\protect\citeauthoryear{{Kipping}, {Schmitt}, {Huang}, {Torres},
  {Nesvorn{\'y}}, {Buchhave}, {Hartman}  \& {Bakos}}{{Kipping}
  et~al.}{2015}]{Kipping2015}
{Kipping} D.~M.,  {Schmitt} A.~R.,  {Huang} X.,  {Torres} G.,  {Nesvorn{\'y}}
  D.,  {Buchhave} L.~A.,  {Hartman} J.,   {Bakos} G.~{\'A}.,  2015, \mn@doi
  [The Astrophysical Journal] {10.1088/0004-637X/813/1/14}, \href
  {https://ui.adsabs.harvard.edu/abs/2015ApJ...813...14K} {813, 14}

\bibitem[\protect\citeauthoryear{{Kov{\'a}cs}, {Zucker}  \&
  {Mazeh}}{{Kov{\'a}cs} et~al.}{2002}]{bls2002}
{Kov{\'a}cs} G.,  {Zucker} S.,   {Mazeh} T.,  2002, \mn@doi [\aap]
  {10.1051/0004-6361:20020802}, \href
  {https://ui.adsabs.harvard.edu/abs/2002A&A...391..369K} {391, 369}

\bibitem[\protect\citeauthoryear{{Kozai}}{{Kozai}}{1962}]{Kozai1962}
{Kozai} Y.,  1962, \mn@doi [The Astronomical Journal] {10.1086/108790}, \href
  {https://ui.adsabs.harvard.edu/abs/1962AJ.....67..591K} {67, 591}

\bibitem[\protect\citeauthoryear{{Kreidberg}}{{Kreidberg}}{2015}]{Kreidberg15}
{Kreidberg} L.,  2015, \mn@doi [\pasp] {10.1086/683602}, \href
  {https://ui.adsabs.harvard.edu/abs/2015PASP..127.1161K} {127, 1161}

\bibitem[\protect\citeauthoryear{{Kurucz}}{{Kurucz}}{2013}]{Kurucz2013}
{Kurucz} R.~L.,  2013, {ATLAS12} (\mn@eprint {ascl} {1303.024})

\bibitem[\protect\citeauthoryear{{Lanza} \& {Mathis}}{{Lanza} \&
  {Mathis}}{2016}]{Lanza2016}
{Lanza} A.~F.,  {Mathis} S.,  2016, \mn@doi [Celestial Mechanics and Dynamical
  Astronomy] {10.1007/s10569-016-9714-z}, \href
  {https://ui.adsabs.harvard.edu/abs/2016CeMDA.126..249L} {126, 249}

\bibitem[\protect\citeauthoryear{{Li}, {Tenenbaum}, {Twicken}, {Burke},
  {Jenkins}, {Quintana}, {Rowe}  \& {Seader}}{{Li} et~al.}{2019}]{Li2019}
{Li} J.,  {Tenenbaum} P.,  {Twicken} J.~D.,  {Burke} C.~J.,  {Jenkins} J.~M.,
  {Quintana} E.~V.,  {Rowe} J.~F.,   {Seader} S.~E.,  2019, \mn@doi [\pasp]
  {10.1088/1538-3873/aaf44d}, \href
  {https://ui.adsabs.harvard.edu/abs/2019PASP..131b4506L} {131, 024506}

\bibitem[\protect\citeauthoryear{{Lidov}}{{Lidov}}{1962}]{Lidov1962}
{Lidov} M.~L.,  1962, \mn@doi [Planetary and Space Science]
  {10.1016/0032-0633(62)90129-0}, \href
  {https://ui.adsabs.harvard.edu/abs/1962P&SS....9..719L} {9, 719}

\bibitem[\protect\citeauthoryear{{Lintott} et~al.,}{{Lintott}
  et~al.}{2008}]{lintott08}
{Lintott} C.~J.,  et~al., 2008, \mn@doi [\mnras]
  {10.1111/j.1365-2966.2008.13689.x}, \href
  {https://ui.adsabs.harvard.edu/abs/2008MNRAS.389.1179L} {389, 1179}

\bibitem[\protect\citeauthoryear{{Lintott} et~al.,}{{Lintott}
  et~al.}{2011}]{lintott11}
{Lintott} C.,  et~al., 2011, VizieR Online Data Catalog, \href
  {https://ui.adsabs.harvard.edu/abs/2011yCat..74100166L} {p. J/MNRAS/410/166}

\bibitem[\protect\citeauthoryear{{Lintott} et~al.,}{{Lintott}
  et~al.}{2013}]{lintott13}
{Lintott} C.~J.,  et~al., 2013, \mn@doi [\aj] {10.1088/0004-6256/145/6/151},
  \href {https://ui.adsabs.harvard.edu/abs/2013AJ....145..151L} {145, 151}

\bibitem[\protect\citeauthoryear{{Livingston} et~al.,}{{Livingston}
  et~al.}{2018}]{Livingston2018}
{Livingston} J.~H.,  et~al., 2018, \mn@doi [\aj] {10.3847/1538-3881/aae778},
  \href {https://ui.adsabs.harvard.edu/abs/2018AJ....156..277L} {156, 277}

\bibitem[\protect\citeauthoryear{{Lomb}}{{Lomb}}{1976}]{Lomb76}
{Lomb} N.~R.,  1976, \mn@doi [\apss] {10.1007/BF00648343}, \href
  {https://ui.adsabs.harvard.edu/abs/1976Ap&SS..39..447L} {39, 447}

\bibitem[\protect\citeauthoryear{Luhn, Bastien, Wright, Johnson, Howard  \&
  Isaacson}{Luhn et~al.}{2019}]{luhn19}
Luhn J.~K.,  Bastien F.~A.,  Wright J.~T.,  Johnson J.~A.,  Howard A.~W.,
  Isaacson H.,  2019, The Astronomical Journal, 157, 149

\bibitem[\protect\citeauthoryear{{Matson}, {Howell}  \& {Ciardi}}{{Matson}
  et~al.}{2019}]{Matson2019}
{Matson} R.~A.,  {Howell} S.~B.,   {Ciardi} D.~R.,  2019, \mn@doi [\aj]
  {10.3847/1538-3881/ab1755}, \href
  {https://ui.adsabs.harvard.edu/abs/2019AJ....157..211M} {157, 211}

\bibitem[\protect\citeauthoryear{{Mayor} et~al.,}{{Mayor}
  et~al.}{2003}]{Mayor2003}
{Mayor} M.,  et~al., 2003, The Messenger, \href
  {https://ui.adsabs.harvard.edu/abs/2003Msngr.114...20M} {114, 20}

\bibitem[\protect\citeauthoryear{{McLaughlin}}{{McLaughlin}}{1924}]{McLaughlin24}
{McLaughlin} D.~B.,  1924, \mn@doi [\apj] {10.1086/142826}, \href
  {https://ui.adsabs.harvard.edu/abs/1924ApJ....60...22M} {60, 22}

\bibitem[\protect\citeauthoryear{{Montet} et~al.,}{{Montet}
  et~al.}{2015}]{Montet2015}
{Montet} B.~T.,  et~al., 2015, \mn@doi [\apj] {10.1088/0004-637X/809/1/25},
  \href {https://ui.adsabs.harvard.edu/abs/2015ApJ...809...25M} {809, 25}

\bibitem[\protect\citeauthoryear{{Morton}}{{Morton}}{2012}]{morton12}
{Morton} T.~D.,  2012, \mn@doi [\apj] {10.1088/0004-637X/761/1/6}, \href
  {http://adsabs.harvard.edu/abs/2012ApJ...761....6M} {761, 6}

\bibitem[\protect\citeauthoryear{{Morton}}{{Morton}}{2015}]{Morton2015}
{Morton} T.~D.,  2015, {VESPA: False positive probabilities calculator},
  Astrophysics Source Code Library (\mn@eprint {ascl} {1503.011})

\bibitem[\protect\citeauthoryear{{Morton}, {Bryson}, {Coughlin}, {Rowe},
  {Ravichandran}, {Petigura}, {Haas}  \& {Batalha}}{{Morton}
  et~al.}{2016}]{Morton2016}
{Morton} T.~D.,  {Bryson} S.~T.,  {Coughlin} J.~L.,  {Rowe} J.~F.,
  {Ravichandran} G.,  {Petigura} E.~A.,  {Haas} M.~R.,   {Batalha} N.~M.,
  2016, \mn@doi [\apj] {10.3847/0004-637X/822/2/86}, \href
  {https://ui.adsabs.harvard.edu/abs/2016ApJ...822...86M} {822, 86}

\bibitem[\protect\citeauthoryear{{Munari} et~al.,}{{Munari}
  et~al.}{2014}]{Munari2014}
{Munari} U.,  et~al., 2014, \mn@doi [\aj] {10.1088/0004-6256/148/5/81}, \href
  {https://ui.adsabs.harvard.edu/abs/2014AJ....148...81M} {148, 81}

\bibitem[\protect\citeauthoryear{{Pearson}, {Palafox}  \& {Griffith}}{{Pearson}
  et~al.}{2018}]{Pearson2018}
{Pearson} K.~A.,  {Palafox} L.,   {Griffith} C.~A.,  2018, \mn@doi [\mnras]
  {10.1093/mnras/stx2761}, \href
  {https://ui.adsabs.harvard.edu/abs/2018MNRAS.474..478P} {474, 478}

\bibitem[\protect\citeauthoryear{{Pecaut} \& {Mamajek}}{{Pecaut} \&
  {Mamajek}}{2013}]{Pecaut2013}
{Pecaut} M.~J.,  {Mamajek} E.~E.,  2013, \mn@doi [The Astrophysical Journal
  Supplement Series] {10.1088/0067-0049/208/1/9}, \href
  {https://ui.adsabs.harvard.edu/abs/2013ApJS..208....9P} {208, 9}

\bibitem[\protect\citeauthoryear{{Persson} et~al.,}{{Persson}
  et~al.}{2018}]{Persson2018}
{Persson} C.~M.,  et~al., 2018, \mn@doi [\aap] {10.1051/0004-6361/201832867},
  \href {https://ui.adsabs.harvard.edu/abs/2018A&A...618A..33P} {618, A33}

\bibitem[\protect\citeauthoryear{{Piskunov} \& {Valenti}}{{Piskunov} \&
  {Valenti}}{2017}]{Piskunov2017}
{Piskunov} N.,  {Valenti} J.~A.,  2017, \mn@doi [\aap]
  {10.1051/0004-6361/201629124}, \href
  {https://ui.adsabs.harvard.edu/abs/2017A&A...597A..16P} {597, A16}

\bibitem[\protect\citeauthoryear{{Rappaport} et~al.,}{{Rappaport}
  et~al.}{2019}]{Rappaport2019}
{Rappaport} S.,  et~al., 2019, \mn@doi [Monthly Notices of the Royal
  Astronomical Society] {10.1093/mnras/stz1772}, \href
  {https://ui.adsabs.harvard.edu/abs/2019MNRAS.488.2455R} {488, 2455}

\bibitem[\protect\citeauthoryear{{Ricker} et~al.,}{{Ricker}
  et~al.}{2015}]{ricker15}
{Ricker} G.~R.,  et~al., 2015, \mn@doi [Journal of Astronomical Telescopes,
  Instruments, and Systems] {10.1117/1.JATIS.1.1.014003}, \href
  {https://ui.adsabs.harvard.edu/abs/2015JATIS...1a4003R} {1, 014003}

\bibitem[\protect\citeauthoryear{{Rossiter}}{{Rossiter}}{1924}]{Rossiter24}
{Rossiter} R.~A.,  1924, \mn@doi [\apj] {10.1086/142825}, \href
  {https://ui.adsabs.harvard.edu/abs/1924ApJ....60...15R} {60, 15}

\bibitem[\protect\citeauthoryear{{Rowe} et~al.,}{{Rowe} et~al.}{2014}]{rowe14}
{Rowe} J.~F.,  et~al., 2014, \mn@doi [\apj] {10.1088/0004-637X/784/1/45}, \href
  {https://ui.adsabs.harvard.edu/abs/2014ApJ...784...45R} {784, 45}

\bibitem[\protect\citeauthoryear{{Rowe} et~al.,}{{Rowe} et~al.}{2015}]{rowe15}
{Rowe} J.~F.,  et~al., 2015, \mn@doi [\apjs] {10.1088/0067-0049/217/1/16},
  \href {https://ui.adsabs.harvard.edu/abs/2015ApJS..217...16R} {217, 16}

\bibitem[\protect\citeauthoryear{{Ryabchikova}, {Piskunov}, {Kurucz},
  {Stempels}, {Heiter}, {Pakhomov}  \& {Barklem}}{{Ryabchikova}
  et~al.}{2015}]{Ryabchikova2015}
{Ryabchikova} T.,  {Piskunov} N.,  {Kurucz} R.~L.,  {Stempels} H.~C.,  {Heiter}
  U.,  {Pakhomov} Y.,   {Barklem} P.~S.,  2015, \mn@doi [\physscr]
  {10.1088/0031-8949/90/5/054005}, \href
  {https://ui.adsabs.harvard.edu/abs/2015PhyS...90e4005R} {90, 054005}

\bibitem[\protect\citeauthoryear{{Santerne} et~al.,}{{Santerne}
  et~al.}{2015}]{santerne15}
{Santerne} A.,  et~al., 2015, \mn@doi [\mnras] {10.1093/mnras/stv1080}, \href
  {https://ui.adsabs.harvard.edu/abs/2015MNRAS.451.2337S} {451, 2337}

\bibitem[\protect\citeauthoryear{{Scargle}}{{Scargle}}{1982}]{Scargle82}
{Scargle} J.~D.,  1982, \mn@doi [\apj] {10.1086/160554}, \href
  {https://ui.adsabs.harvard.edu/abs/1982ApJ...263..835S} {263, 835}

\bibitem[\protect\citeauthoryear{{Schlegel}, {Finkbeiner}  \&
  {Davis}}{{Schlegel} et~al.}{1998}]{Schlegel98}
{Schlegel} D.~J.,  {Finkbeiner} D.~P.,   {Davis} M.,  1998, \mn@doi [\apj]
  {10.1086/305772}, \href
  {https://ui.adsabs.harvard.edu/abs/1998ApJ...500..525S} {500, 525}

\bibitem[\protect\citeauthoryear{{Schmitt} et~al.,}{{Schmitt}
  et~al.}{2014a}]{schmitt14a}
{Schmitt} J.~R.,  et~al., 2014a, \mn@doi [\aj] {10.1088/0004-6256/148/2/28},
  \href {https://ui.adsabs.harvard.edu/abs/2014AJ....148...28S} {148, 28}

\bibitem[\protect\citeauthoryear{{Schmitt} et~al.,}{{Schmitt}
  et~al.}{2014b}]{schmitt14b}
{Schmitt} J.~R.,  et~al., 2014b, \mn@doi [\apj] {10.1088/0004-637X/795/2/167},
  \href {https://ui.adsabs.harvard.edu/abs/2014ApJ...795..167S} {795, 167}

\bibitem[\protect\citeauthoryear{Schneider}{Schneider}{2000}]{Schneider2000}
Schneider J.,  2000, in From Giant Planets to Cool Stars. p.~284

\bibitem[\protect\citeauthoryear{{Schwamb} et~al.,}{{Schwamb}
  et~al.}{2012}]{schwamb12}
{Schwamb} M.~E.,  et~al., 2012, \mn@doi [\apj] {10.1088/0004-637X/754/2/129},
  \href {https://ui.adsabs.harvard.edu/abs/2012ApJ...754..129S} {754, 129}

\bibitem[\protect\citeauthoryear{{Schwamb} et~al.,}{{Schwamb}
  et~al.}{2013}]{schwamb13}
{Schwamb} M.~E.,  et~al., 2013, \mn@doi [\apj] {10.1088/0004-637X/768/2/127},
  \href {https://ui.adsabs.harvard.edu/abs/2013ApJ...768..127S} {768, 127}

\bibitem[\protect\citeauthoryear{{Smith} et~al.,}{{Smith}
  et~al.}{2012}]{Smith2012}
{Smith} J.~C.,  et~al., 2012, \mn@doi [\pasp] {10.1086/667697}, \href
  {https://ui.adsabs.harvard.edu/abs/2012PASP..124.1000S} {124, 1000}

\bibitem[\protect\citeauthoryear{{Southworth}}{{Southworth}}{2011}]{tepcat}
{Southworth} J.,  2011, \mn@doi [\mnras] {10.1111/j.1365-2966.2011.19399.x},
  \href {https://ui.adsabs.harvard.edu/abs/2011MNRAS.417.2166S} {417, 2166}

\bibitem[\protect\citeauthoryear{{Stassun} \& {Torres}}{{Stassun} \&
  {Torres}}{2016}]{Stassun16}
{Stassun} K.~G.,  {Torres} G.,  2016, \mn@doi [\aj]
  {10.3847/0004-6256/152/6/180}, \href
  {https://ui.adsabs.harvard.edu/abs/2016AJ....152..180S} {152, 180}

\bibitem[\protect\citeauthoryear{{Stassun} \& {Torres}}{{Stassun} \&
  {Torres}}{2018}]{StassunTorres18}
{Stassun} K.~G.,  {Torres} G.,  2018, \mn@doi [\apj]
  {10.3847/1538-4357/aacafc}, \href
  {https://ui.adsabs.harvard.edu/abs/2018ApJ...862...61S} {862, 61}

\bibitem[\protect\citeauthoryear{{Stassun}, {Collins}  \& {Gaudi}}{{Stassun}
  et~al.}{2017}]{Stassun17}
{Stassun} K.~G.,  {Collins} K.~A.,   {Gaudi} B.~S.,  2017, \mn@doi [\aj]
  {10.3847/1538-3881/aa5df3}, \href
  {https://ui.adsabs.harvard.edu/abs/2017AJ....153..136S} {153, 136}

\bibitem[\protect\citeauthoryear{{Stassun}, {Corsaro}, {Pepper}  \&
  {Gaudi}}{{Stassun} et~al.}{2018}]{Stassun18}
{Stassun} K.~G.,  {Corsaro} E.,  {Pepper} J.~A.,   {Gaudi} B.~S.,  2018,
  \mn@doi [\aj] {10.3847/1538-3881/aa998a}, \href
  {https://ui.adsabs.harvard.edu/abs/2018AJ....155...22S} {155, 22}

\bibitem[\protect\citeauthoryear{{Stassun} et~al.,}{{Stassun}
  et~al.}{2019}]{Stassun19}
{Stassun} K.~G.,  et~al., 2019, arXiv e-prints, \href
  {https://ui.adsabs.harvard.edu/abs/2019arXiv190510694S} {p. arXiv:1905.10694}

\bibitem[\protect\citeauthoryear{{Storch}, {Lai}  \& {Anderson}}{{Storch}
  et~al.}{2017}]{Storch2017}
{Storch} N.~I.,  {Lai} D.,   {Anderson} K.~R.,  2017, \mn@doi [Monthly Notices
  of the Royal Astronomical Society] {10.1093/mnras/stw3018}, \href
  {https://ui.adsabs.harvard.edu/abs/2017MNRAS.465.3927S} {465, 3927}

\bibitem[\protect\citeauthoryear{{Stumpe} et~al.,}{{Stumpe}
  et~al.}{2012}]{Stumpe2012}
{Stumpe} M.~C.,  et~al., 2012, \mn@doi [\pasp] {10.1086/667698}, \href
  {https://ui.adsabs.harvard.edu/abs/2012PASP..124..985S} {124, 985}

\bibitem[\protect\citeauthoryear{{Tokovinin}}{{Tokovinin}}{2018}]{Tokovinin2018}
{Tokovinin} A.,  2018, \mn@doi [\pasp] {10.1088/1538-3873/aaa7d9}, \href
  {https://ui.adsabs.harvard.edu/abs/2018PASP..130c5002T} {130, 035002}

\bibitem[\protect\citeauthoryear{{Tokovinin}, {Fischer}, {Bonati}, {Giguere},
  {Moore}, {Schwab}, {Spronck}  \& {Szymkowiak}}{{Tokovinin}
  et~al.}{2013}]{Tokovinin2013}
{Tokovinin} A.,  {Fischer} D.~A.,  {Bonati} M.,  {Giguere} M.~J.,  {Moore} P.,
  {Schwab} C.,  {Spronck} J. F.~P.,   {Szymkowiak} A.,  2013, \mn@doi [\pasp]
  {10.1086/674012}, \href
  {https://ui.adsabs.harvard.edu/abs/2013PASP..125.1336T} {125, 1336}

\bibitem[\protect\citeauthoryear{Torres, Andersen  \& Gim{\'e}nez}{Torres
  et~al.}{2010a}]{Torres10}
Torres G.,  Andersen J.,   Gim{\'e}nez A.,  2010a, The Astronomy and
  Astrophysics Review, 18, 67

\bibitem[\protect\citeauthoryear{{Torres}, {Andersen}  \&
  {Gim{\'e}nez}}{{Torres} et~al.}{2010b}]{Torres2010}
{Torres} G.,  {Andersen} J.,   {Gim{\'e}nez} A.,  2010b, \mn@doi [\aapr]
  {10.1007/s00159-009-0025-1}, \href
  {https://ui.adsabs.harvard.edu/abs/2010A&ARv..18...67T} {18, 67}

\bibitem[\protect\citeauthoryear{{Twicken} et~al.,}{{Twicken}
  et~al.}{2018}]{Twicken2018}
{Twicken} J.~D.,  et~al., 2018, \mn@doi [\pasp] {10.1088/1538-3873/aab694},
  \href {https://ui.adsabs.harvard.edu/abs/2018PASP..130f4502T} {130, 064502}

\bibitem[\protect\citeauthoryear{{Valenti} \& {Piskunov}}{{Valenti} \&
  {Piskunov}}{1996}]{Valenti1996}
{Valenti} J.~A.,  {Piskunov} N.,  1996, \aaps, \href
  {https://ui.adsabs.harvard.edu/abs/1996A&AS..118..595V} {118, 595}

\bibitem[\protect\citeauthoryear{{Van Eylen} et~al.,}{{Van Eylen}
  et~al.}{2019}]{VanEylen2019}
{Van Eylen} V.,  et~al., 2019, \mn@doi [\aj] {10.3847/1538-3881/aaf22f}, \href
  {https://ui.adsabs.harvard.edu/abs/2019AJ....157...61V} {157, 61}

\bibitem[\protect\citeauthoryear{{Vanderburg} et~al.,}{{Vanderburg}
  et~al.}{2019}]{Vanderburg2019}
{Vanderburg} A.,  et~al., 2019, \mn@doi [\apjl] {10.3847/2041-8213/ab322d},
  \href {https://ui.adsabs.harvard.edu/abs/2019ApJ...881L..19V} {881, L19}

\bibitem[\protect\citeauthoryear{{Vazan}, {Kovetz}, {Podolak}  \&
  {Helled}}{{Vazan} et~al.}{2013}]{Vazan2013}
{Vazan} A.,  {Kovetz} A.,  {Podolak} M.,   {Helled} R.,  2013, \mn@doi [\mnras]
  {10.1093/mnras/stt1248}, \href
  {https://ui.adsabs.harvard.edu/abs/2013MNRAS.434.3283V} {434, 3283}

\bibitem[\protect\citeauthoryear{{Wang} et~al.,}{{Wang} et~al.}{2013}]{wang13}
{Wang} J.,  et~al., 2013, \mn@doi [\apj] {10.1088/0004-637X/776/1/10}, \href
  {https://ui.adsabs.harvard.edu/abs/2013ApJ...776...10W} {776, 10}

\bibitem[\protect\citeauthoryear{{Wang} et~al.,}{{Wang} et~al.}{2015}]{wang15}
{Wang} J.,  et~al., 2015, \mn@doi [\apj] {10.1088/0004-637X/815/2/127}, \href
  {https://ui.adsabs.harvard.edu/abs/2015ApJ...815..127W} {815, 127}

\bibitem[\protect\citeauthoryear{{Winn}}{{Winn}}{2010}]{Winn2010}
{Winn} J.~N.,  2010, {Exoplanet Transits and Occultations}.
University of Arizona Press, pp 55--77

\bibitem[\protect\citeauthoryear{{Yee}, {Petigura}  \& {von Braun}}{{Yee}
  et~al.}{2017}]{Yee2017}
{Yee} S.~W.,  {Petigura} E.~A.,   {von Braun} K.,  2017, \mn@doi [\apj]
  {10.3847/1538-4357/836/1/77}, \href
  {https://ui.adsabs.harvard.edu/abs/2017ApJ...836...77Y} {836, 77}

\bibitem[\protect\citeauthoryear{{Ziegler}, {Tokovinin}, {Briceno}, {Mang},
  {Law}  \& {Mann}}{{Ziegler} et~al.}{2019}]{Ziegler2019}
{Ziegler} C.,  {Tokovinin} A.,  {Briceno} C.,  {Mang} J.,  {Law} N.,   {Mann}
  A.~W.,  2019, arXiv e-prints, \href
  {https://ui.adsabs.harvard.edu/abs/2019arXiv190810871Z} {p. arXiv:1908.10871}

\bibitem[\protect\citeauthoryear{{Zink} et~al.,}{{Zink}
  et~al.}{2019}]{Zink2019}
{Zink} J.~K.,  et~al., 2019, \mn@doi [Research Notes of the American
  Astronomical Society] {10.3847/2515-5172/ab0a02}, \href
  {https://ui.adsabs.harvard.edu/abs/2019RNAAS...3b..43Z} {3, 43}

\bibitem[\protect\citeauthoryear{{Zucker} \& {Giryes}}{{Zucker} \&
  {Giryes}}{2018}]{Zucker2018}
{Zucker} S.,  {Giryes} R.,  2018, \mn@doi [\aj] {10.3847/1538-3881/aaae05},
  \href {https://ui.adsabs.harvard.edu/abs/2018AJ....155..147Z} {155, 147}

\makeatother
\end{thebibliography}
\vspace{7mm}
\noindent
\textit{
$^{1}$Department of Physics, University of Oxford, Keble Road, Oxford OX1 3RH, UK\\
$^{2}$Department of Physics and Astronomy, Louisiana State University, Baton Rouge, LA 70803 USA\\
$^{3}$Cerro Tololo Inter-American Observatory, Casilla 603, La Serena, Chile\\
$^{4}$Department of Physics, University of Warwick, Gibbet Hill Road, Coventry, CV4 7AL, UK\\
$^{5}$Centre for Exoplanets and Habitability, University of Warwick, Gibbet Hill road, Coventry, CV4 7AL, UK\\
$^{6}$NASA Ames Research Center, Moffett Field, CA 94035, USA\\
$^{7}$Department of Astronomy and Astrophysics, University of
Chicago, 5640 S. Ellis Ave, Chicago, IL 60637, USA\\
$^{8}$Departamento de Astronom\'ia, Universidad de Guanajuato, Callej\'on de Jalisco S/N, Col. Valenciana CP, 36023 Guanajuato, Gto, M\'exico \\
$^{9}$Chalmers University of Technology, Department of Space, Earth and Environment, Onsala Space Observatory, SE-439 92 Onsala, Sweden\\
$^{10}$Leiden Observatory, University of Leiden, PO Box 9513, 2300 RA, Leiden, The Netherlands\\
$^{11}$Dipartimento di Fisica, Universit\'a di Torino, Via P. Giuria 1, I-10125, Torino, Italy\\
$^{12}$Research School of Astronomy \& Astrophysics, Mount Stromlo Observatory, Australian National University, Cotter Road, Weston Creek, ACT 2611, Australia\\
$^{13}$Department of Physics, and Kavli Institute for Astrophysics and Space Research, Massachusetts Institute of Technology, Cambridge, MA 02139, USA \\
$^{14}$DTU Space, National Space Institute, Technical University of Denmark, Elektrovej 327, DK-2800 Lyngby, Denmark\\
$^{15}$Department of Astronomy, The University of Texas at Austin, Austin, TX 78712, USA\\
$^{16}$Department of Physics and Astronomy, The University of North Carolina at Chapel Hill, Chapel Hill, NC 27599-3255, USA\\
$^{17}$Australian Astronomical Optics, 105 Delhi Rd, North Ryde, NSW 2113, Australia\\
$^{18}$Department of Physics and Astronomy, Macquarie University, NSW 2109, Australia\\
$^{19}$Gemini Observatory, Northern Operations Center, 670 North A'ohoku Place, Hilo, HI 96720, USA\\
$^{20}$Astrophysics Research Centre, Queen's University Belfast, Belfast BT7 1NN, UK\\
$^{21}$Vanderbilt University, Department of Physics \& Astronomy, 6301 Stevenson Center Ln., Nashville, TN 37235, USA\\
$^{22}$Fisk University, Department of Physics, 1000 17th Ave. N., Nashville, TN 37208, USA\\
$^{23}$Aix Marseille Univ, CNRS, CNES, Laboratoire d’Astrophysique de Marseille, France\\
$^{24}$Department of Astronomy, Yale University, New Haven, CT 06511, USA\\
$^{25}$Sydney Institute for Astronomy, School of Physics, University of Sydney, NSW 2006, Australia\\
$^{26}$CSIRO Astronomy and Space Science, PO Box 76, Epping, NSW 1710, Australia\\
$^{27}$Dunlap Institute for Astronomy and Astrophysics, University of Toronto, 50 St. George Street, Toronto, Ontario M5S 3H4, Canada\\
$^{28}$Citizen Scientist, Zooniverse c/o University of Oxford, Keble Road, Oxford OX1 3RH, UK\\
}

%%%%%%%%%%%%%%%%%%%%%%%%%%%%%%%%%%%%%%%%%%%%%%%%%%

%%%%%%%%%%%%%%%%% APPENDICES %%%%%%%%%%%%%%%%%%%%%

%\appendix



%%%%%%%%%%%%%%%%%%%%%%%%%%%%%%%%%%%%%%%%%%%%%%%%%%

% Don't change these lines
\bsp	% typesetting comment
\label{lastpage}
\end{document}